\documentclass[12pt,showpacs,showkeys,prb]{revtex4}

\usepackage{amssymb}
\usepackage{amsmath}
\usepackage{graphicx}
\usepackage{amsbsy}
\usepackage{color}

\newcommand{\bq}{\begin{eqnarray}}
\newcommand{\eq}{\end{eqnarray}}
\newcommand{\bqn}{\begin{eqnarray*}}
\newcommand{\eqn}{\end{eqnarray*}}
\newcommand{\rrr}{\mathbf{r}}
\newcommand{\kkk}{\mathbf{k}}
\newcommand{\sss}{\mathbf{s}}

\begin{document}
%%%%%%%%%%%%%%%%%%%%%%%%%%%%%%%%%%%%%%%%%%%%%%%%%%%%%%%%%%%%%%%%%%%%%%%%%%%%%%%
%%%%%%%%%%%%%%%%%%%%%%%%%%%%%%%%%%%%%%%%%%%%%%%%%%%%%%%%%%%%%%%%%%%%%%%%%%%%%%%
%%%%%%%%%%%%%%%%%%%%%%%%%%%%%%%%%%%%%%%%%%%%%%%%%%%%%%%%%%%%%%%%%%%%%%%%%%%%%%%
\title{Patchy sticky hard spheres: analytical study and Monte Carlo simulations}

\author{Riccardo Fantoni}
\email{rfantoni@unive.it}
\author{Domenico Gazzillo}
\email{gazzillo@unive.it}
\author{Achille Giacometti}
\email{achille@unive.it}
\affiliation{Dipartimento di Chimica Fisica, Universit\`a di Venezia, S. Marta DD
2137, I-30123 Venezia, Italy} 
\author{Mark A. Miller}
\email{mam1000@cam.ac.uk}
\affiliation{University Chemical Laboratory, Lensfield Road, Cambridge
CB2 1EW, United Kingdom}
\author{Giorgio Pastore}
\email{pastore@ts.infn.it}
\affiliation{Dipartimento di Fisica Teorica, Universit\`a di Trieste,
Strada Costiera 11, 34100 Trieste, Italy}
\date{\today}

\begin{abstract}
We consider a fluid of hard spheres bearing one or two uniform circular adhesive
patches, distributed so as not to overlap.
Two spheres interact via a ``sticky'' Baxter potential
if the line joining the centers of the two spheres intersects
a patch on each sphere, and via a hard sphere potential otherwise.
We analyze the location of the fluid-fluid transition and of the percolation
line as a function of the size of the patch (the fractional coverage of the sphere's
surface) and of the number of patches within a virial expansion up to
third order and within the first two terms (C0 and C1) of a class of closures Cn
hinging on a density expansion of the direct correlation function.
We find that the locations of the two lines depend sensitively
on both the total adhesive coverage and its distribution. The treatment is almost
fully analytical within the chosen approximate theory.
We test our findings by means of specialized Monte Carlo (MC) simulations and
find the main qualitative features of the critical behaviour to be well captured 
in spite of the low density perturbative nature of the closure.
The introduction of anisotropic attractions into a model suspension of spherical
particles is a first step towards a more realistic description of globular proteins
in solution.
\end{abstract}

\maketitle

%%%%%%%%%%%%%%%%%%%%%%%%%%%%%%%%%%%%%%%%%%%%%%%%%%%%%%%%%%%%%%%%%%%%%%%%%%%%%%%
\section{Introduction}
%%%%%%%%%%%%%%%%%%%%%%%%%%%%%%%%%%%%%%%%%%%%%%%%%%%%%%%%%%%%%%%%%%%%%%%%%%%%%%%

The idea of modeling fluids as systems of spherical particles with orientationally
dependent attraction dates back at least as far as Boltzmann,
who envisaged chemical attraction between atoms only when ``their sensitive regions
are in contact''.\cite{Boltzmann95}  Models of this type, featuring patchy interactions,
are currently experiencing renewed relevance in the context of colloidal
and biological matter in contrast
to their original conception in connection with fluids of atoms and small molecules.
\cite{Jackson88,Blum90,Busch94,Ghonasgi95,Busch96,Lomakin99,Sear99,Mileva00,Sciortino02,
Starr03,Kern03,Zhang04,Glotzer04,Cho05, Sciortino05,Sciortino06}
\par
The new interest arises for various reasons.
On the technological side, patchy particles give 
the possibility of designing self-assembling
nano-scale devices through anisotropic decorations of the particle surface by
means of organic or biological molecules.\cite{Zhang04,Glotzer04,Cho05}
Nature provides inspiration for what might be achieved in this area, a
particularly elegant example being the self assembly of virus capsids.
These protein shells are monodisperse and highly symmetric, and are composed
of identical subunits.  Simplified
descriptions of icosahedral virus capsids are currently being formulated using
spherical subunits with directional interactions,\cite{Hagan06,Wilber07}
and the possibility of adopting similar schemes to self assemble other
target structures is being explored.\cite{Zhang04}  This level of organization
inevitably requires a certain specificity in the interactions between the subunits
as well as measures to prevent further aggregation of the assembled objects.
\par

Less specific patchy interactions give rise to associating fluids containing a
distribution of cluster sizes or extended gel-like networks.  The key feature of
such systems is a set of point-like sites on the particle surface, leading
to strongly directional bonding with a maximum of one bond per
site.\cite{Blum90,Sear99,Sciortino02,Starr03,Sciortino05,Sciortino06}
This type of interaction has proven invaluable in elucidating
the interplay between fluid--fluid and sol--gel transitions.  One
advantage of these models is that powerful analytical
tools are available for them, such as
Wertheim's thermodynamic perturbation theory,\cite{Wertheim84}
which yields accurate results under experimentally
realistic conditions.\cite{Kalyuzhnyi91,Kalyuzhnyi94}
\par

In contrast to these models with attractive spots, one can envisage particles that
interact through larger attractive regions on their surface, for example, globular
proteins with patches of hydrophobic (nonpolar) amino acids exposed at the surface.
Isotropic potentials have been remarkably successful in modeling
the phase diagrams of certain proteins,\cite{Piazza98,Giacometti05,Pellicane04}
but it seems that not all features of their coexistence curves can be properly
explained by such simple interactions.\cite{Lomakin96}  In this sort of
system, it
seems more appropriate to consider regions with short-range attractive
forces\cite{Jackson88,Busch94,Ghonasgi95,Busch96,Lomakin99,Mileva00,Kern03}
rather than site--site bonds.
These attractive patches are capable of sustaining as many ``bonds'' as permitted by
geometry.  The size of the patch therefore becomes an important new parameter
that does not arise in most work on associating fluids.  We note, however,
that for sufficiently narrow patches, the two models become essentially equivalent.
\par
In the present work, we focus on a simple yet physical model
which is a variation of those treated in.\cite{Kern03}
We consider uniform circular patches distributed on the surface of the sphere
in such a way that they do not overlap.  The
patches are delimited by circles which can be defined by
the associated solid angles.  Two particles experience an adhesive attraction
only when a patch on one sphere touches a patch on the other.
The adhesion is of Baxter's type,\cite{Baxter68} i.e.,
the attraction has infinitesimal range, acting only when the particles are exactly
in contact, as described in the next section.
This model has the advantage that it can be tackled with
analytical tools, unlike most other models for which not even the isotropic
analogue bears this appealing feature.
Various issues arise in this sort of model relating to the stability of the liquid phase
with respect to crystalline solid phases, and these points have been studied in
Refs.~\onlinecite{Stell91,Charbonneau07}.
\par
The integral equation theory of fluids with an angularly dependent pair 
potential is complicated by the fact that the pair distribution function 
is also angularly dependent.\cite{Gray}  In the general case one must appeal
to the symmetries of the fluid (translational invariance,
rotational invariance, invariance under permutation of like particles,
and invariance under the symmetry operations of the individual particles
and of the correlation functions) in order to simplify the
problem.\cite{Blum72a,Blum72b,Blum73}
In some cases, it is possible to factorize the angular dependence of the
Ornstein-Zernike (OZ) equation.  For example, the factorization for a fluid of
dipolar particles has long been known,\cite{Wertheim71} and in Ref.~\onlinecite{Gazzillo07}
it is shown how to solve the dipole-dipole angular distribution of attraction in
the adhesive limit within the Percus-Yevick (PY) framework.  However, the
dipolar case hinges on exploiting a special property of this particular angular
distribution that is particularly useful for the angular convolution in the OZ equation.
In contrast, for an angular dependence with
discontinuities, such as the circular patches treated here, any approach relying
on a spherical harmonic expansion would prove a formidable task
due to the large number of terms necessary to capture the discontinuities.
\par
In the present model we therefore follow a different route based on two
parallel and related schemes. We first
perform a virial expansion up to the third virial coefficient.
We then proceed to study a class of closures (denoted C0, C1,...) which
were proposed in Ref.~\onlinecite{Gazzillo04} and are based on a density
expansion of the direct correlation function. In particular, the zeroth-order term
(C0) turns out to be equivalent to a modified mean spherical approximation,
whereas the first-order (C1) is known to provide the correct third virial
coefficient.\cite{Gazzillo04}
Within both schemes we study the thermodynamics, radial distribution
function and percolation threshold, and compare with specialized Monte Carlo
simulations which were recently devised to this aim.\cite{Miller07}
By varying the size of the adhesive patches and by selecting between one patch and
two diametrically opposite patches, we are able to investigate the roles of both the
total surface coverage and the geometrical distribution of the adhesion.  In both
the one- and two-patch
cases we can change smoothly between small sticky spots, capable of making only one
bond each, and the isotropic adhesive sphere.
We find that 
the position of the critical fluid-fluid transition line and the 
percolation threshold are both
sensitive to the surface coverage. At fixed coverage, there is also
a dependence on the way in which this adhesion is distributed.
\par
Our results can be compared and contrasted with the recent work of Bianchi 
{\it et al.},\cite{Sciortino06} who consider the maximum number of bonds per
particle, rather than the fractional surface coverage, as the key
parameter controlling the location of the critical point. In the
present work we are able to tune both effects, thus illuminating
their specific roles in the location of critical points.

The remainder of the paper is organized as follows.
In section \ref{sect:model} we introduce the model while section \ref{sect:analysis}
contains a description of the analytical and numerical used tools. Results
for the radial distribution function, fluid-fluid transition and percolation threshold
are included in sections \ref{sect:structure}, \ref{sect:coexistence} and 
\ref{sect:percolation}, respectively. Finally, in section \ref{sect:background} the
inclusion of an adhesive background is discussed, and conclusions
and an outlook are contained in section \ref{sect:conclusions}.
  
%%%%%%%%%%%%%%%%%%%%%%%%%%%%%%%%%%%%%%%%%%%%%%%%%%%%%%%%%%%%%%%%%%%%%%%%%%%%%%%
\section{Definition of the model}
\label{sect:model}
%%%%%%%%%%%%%%%%%%%%%%%%%%%%%%%%%%%%%%%%%%%%%%%%%%%%%%%%%%%%%%%%%%%%%%%%%%%%%%%
\subsection{Baxter model with orientationally dependent adhesion}
%%%%%%%%%%%%%%%%%%%%%%%%%%%%%%%%%%%%%%%%%%%%%%%%%%%%%%%%%%%%%%%%%%%%%%%%%%%%%%%

\begin{figure}[ht!]
\begin{center}
\includegraphics[width=10cm]{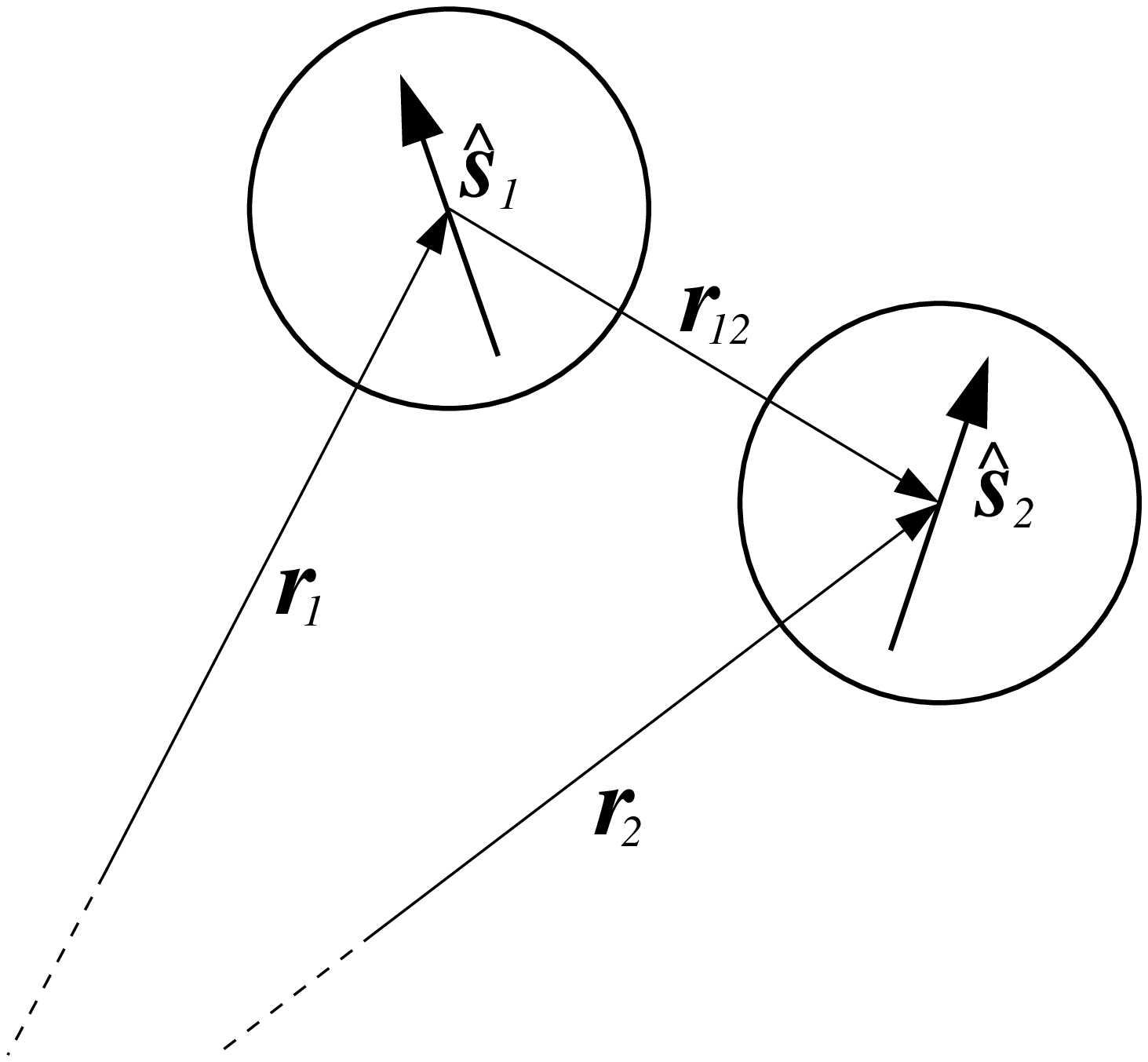}
\end{center}
\caption{}
\label{fig:not}
\end{figure}

We start with some general remarks on the orientational dependence
of a three-dimensional homogeneous fluid of hard spheres with adhesive
pairwise interactions.
Let $\rrr_i$ be the coordinates of the $i^{\rm th}$ particle 
($i=1,2,3,\ldots$) and assume that the
patch distribution on the sphere
has cylindrical symmetry so that
its orientation in space is determined by a unit vector 
$\hat{\sss}_i$ rigidly attached to it. Then 
$\hat{\sss}_i=(\sin\theta_i\cos\varphi_i,\sin\theta_i\sin\varphi_i,
\cos\theta_i)$ 
where $\theta_i$ and $\varphi_i$ are the polar and the azimuthal angles with 
respect to a fixed reference frame (see Fig.~\ref{fig:not}). 
As usual, we introduce the relative coordinates 
$\rrr_{12}=\rrr_2-\rrr_1$ and the associated distance $r_{12}=|\rrr_{12}|$,
and work with the following short-hand notation: 
$(1,2)=(\rrr_{12},\theta_1,\varphi_1,\theta_2,\varphi_2)$ and
$\Omega_i=(\theta_i,\varphi_i)$ for the orientation of
$\hat{\sss}_i$. The orientation of $\hat{\rrr}_{ij}=\rrr_{ij}/r_{ij}$ 
with respect to the same frame of reference will be denoted by
$\Omega_{ij}=(\theta_{ij},\varphi_{ij})$.

The particles interact through a pair potential $\phi(1,2)$,
defined below, which is a generalization of Baxter's sticky hard sphere (SHS)
limit\cite{Baxter68} to orientationally dependent interactions.
We start with
\bq
\beta \Phi(1,2)=\left\{ 
\begin{array}{ll}
+\infty & 0<r<\sigma~, \\ 
\displaystyle-\ln \left[\frac{\epsilon(1,2)}{12\tau}\frac{R}{R
-\sigma}\right] & \sigma\leq r\leq R~, \\ 
0 & r>R~,
\end{array}
\right.  \label{SW}
\eq
where $\beta=1/(k_{B}T)$ ($k_{B}$ being Boltzmann's constant and $T$ the
temperature), $\sigma$ is the diameter of the spheres, and 
$\epsilon(1,2)/\tau$ is a dimensionless adhesion coefficient.  We
define $\phi(1,2)$ through the following limit on the 
Boltzmann factor $e$,
\bq
e(1,2)=\exp[{-\beta\phi(1,2)}]=\lim_{R\to\sigma}\exp[{-\beta\Phi(1,2)}]
=\Theta(r_{12}-\sigma)+\frac{\epsilon(1,2)}{\tau}\frac{\sigma}{12}
\delta(r_{12}-\sigma)~.
\label{SHS}
\eq 
where $\Theta(\cdot)$ is the Heaviside step function and $\delta(\cdot)$ the Dirac
delta function. When $\epsilon(1,2)=1$ we recover the usual Baxter SHS model
and hence the only orientational dependence is included in the definition
of $\epsilon(1,2)$. It is easy to see that $\epsilon(1,2)$
cannot be a simple function of $\hat{\sss}_1$ and $\hat{\sss}_2$ but must also include
a dependence on  
$\hat{\rrr}_{12}=\rrr_{12}/r_{12}$, in order to avoid a trivial corresponding states rescaling.
This point is discussed in Appendix \ref{app:lcs}.
In the present work we shall address
a type of orientational dependence which
was introduced by Kern and Frenkel\cite{Kern03} following a previous
suggestion by Jackson {\it et al.}\cite{Jackson88}

%%%%%%%%%%%%%%%%%%%%%%%%%%%%%%%%%%%%%%%%%%%%%%%%%%%%%%%%%%%%%%%%%%%%%%%%%%%%%%%
\subsection{Patchy sticky hard spheres}
%%%%%%%%%%%%%%%%%%%%%%%%%%%%%%%%%%%%%%%%%%%%%%%%%%%%%%%%%%%%%%%%%%%%%%%%%%%%%%%

\begin{figure}[ht!]
\begin{center}
\includegraphics[width=8cm]{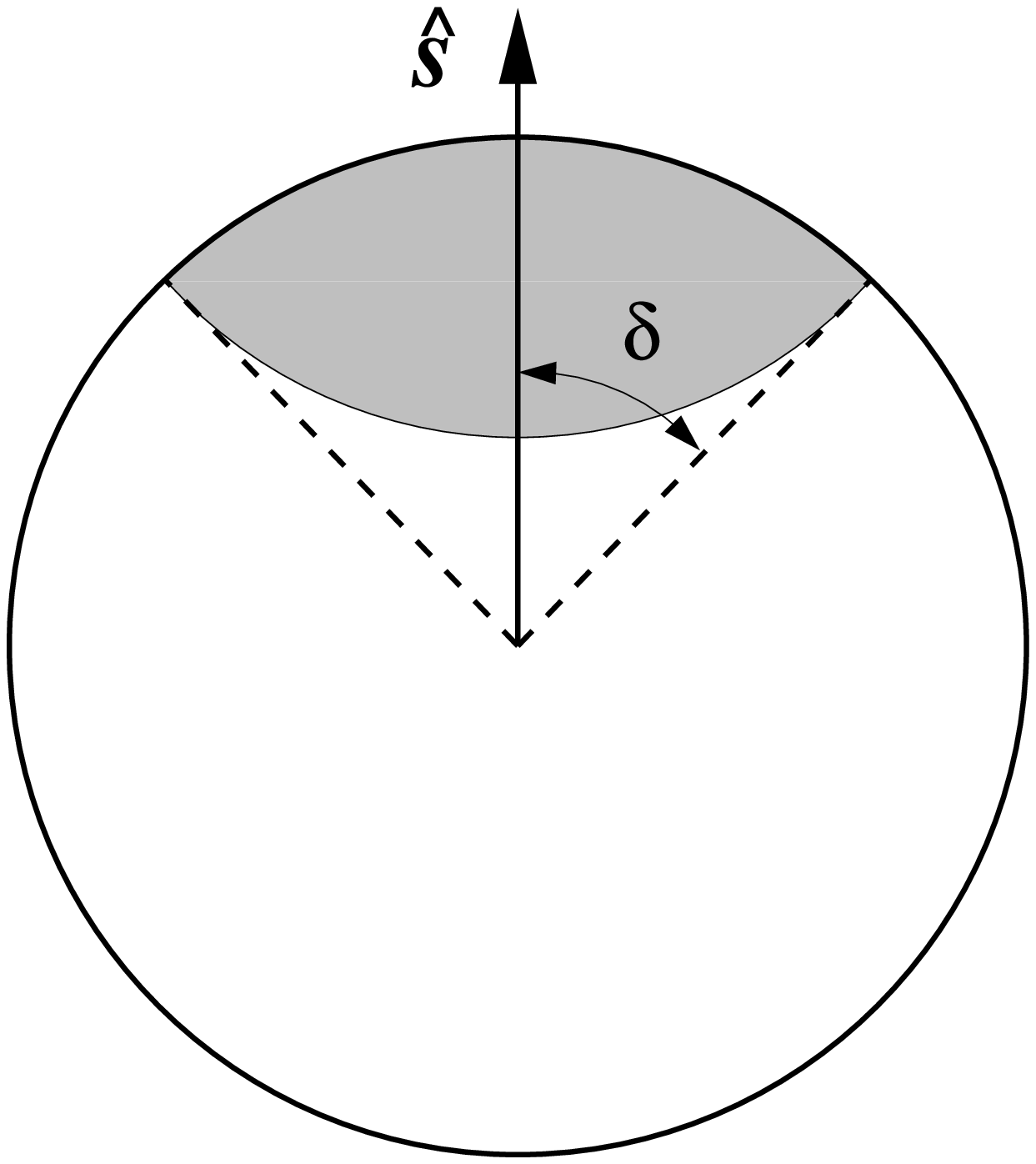}
\includegraphics[width=8cm]{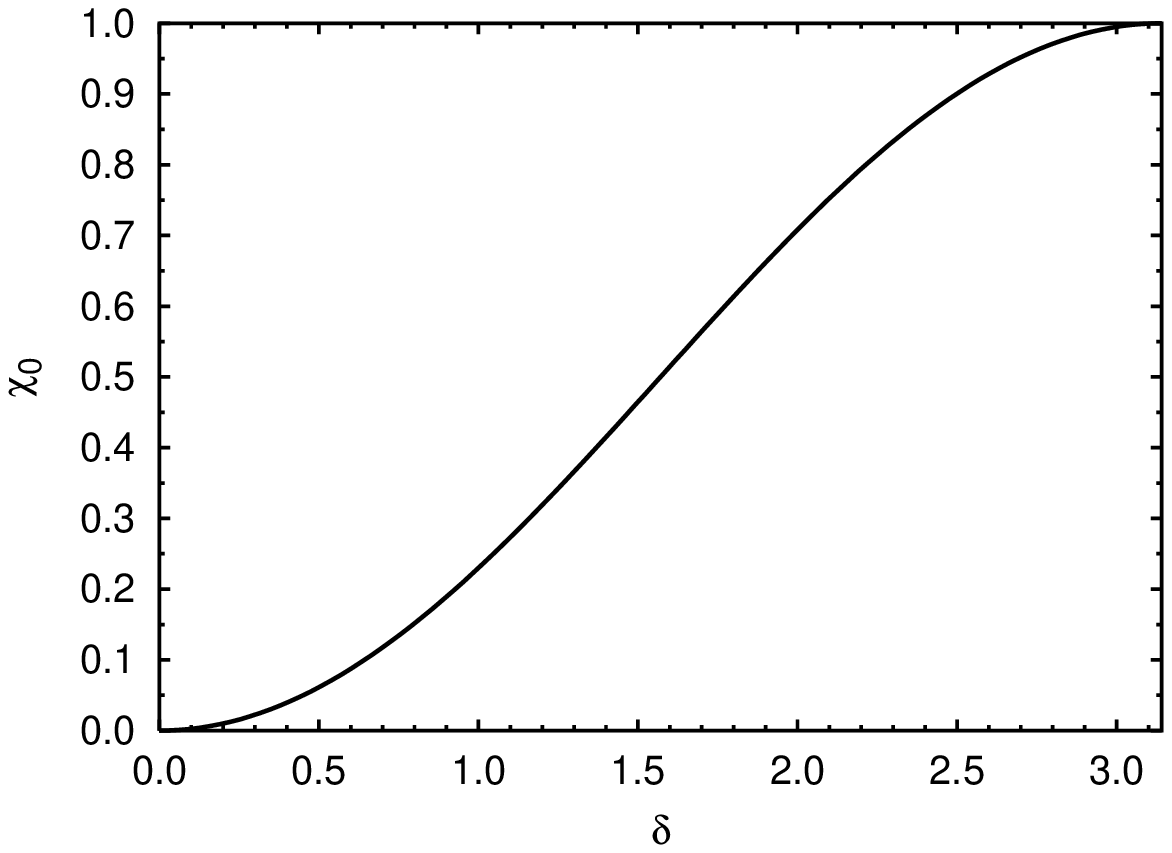}
\end{center}
\caption{}
\label{fig:patch}
\end{figure}

Consider a single hard sphere having one or more identical adhesive circular patches
distributed on its surface in such a way that they do not overlap with one another.
The size of the patch can be specified by the angular amplitude $2 \delta$
as shown in Fig. \ref{fig:patch}.
The unit vector $\hat{\sss}_i^{(p)}$ identifies the direction
from the center of particle $i$ to the center of patch $p$ on the surface 
($p=1,\ldots,n$, the total number of patches).
The sticky area is then given by points $\hat\rrr$ on the
surface of the particle such that the angle between 
$\hat{\sss}_i^{(p)}$ and $\hat{\rrr}$ is smaller than $\delta$. 

\begin{figure}[ht!]
\begin{center}
\includegraphics[width=8cm]{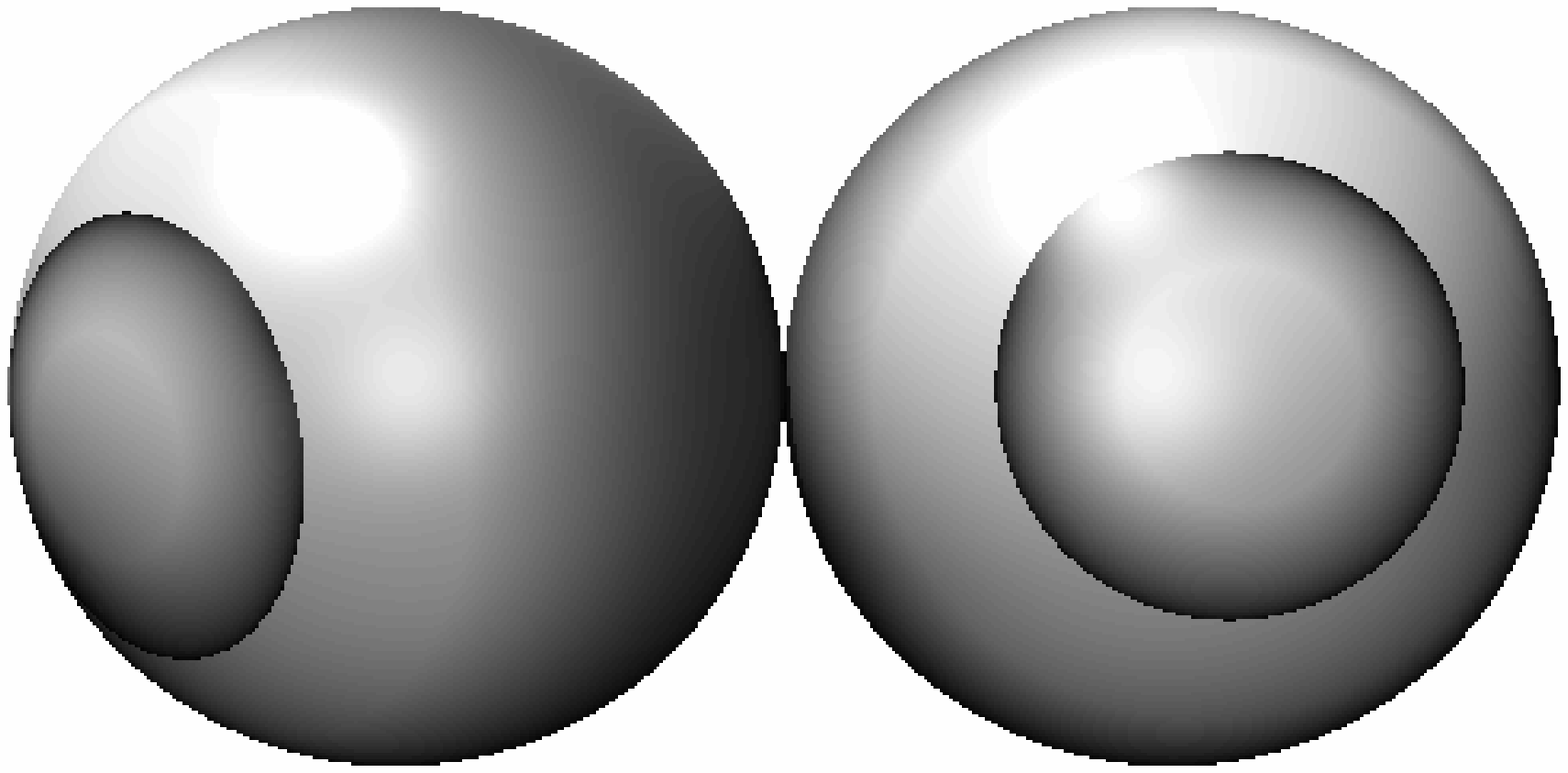}
\includegraphics[width=8cm]{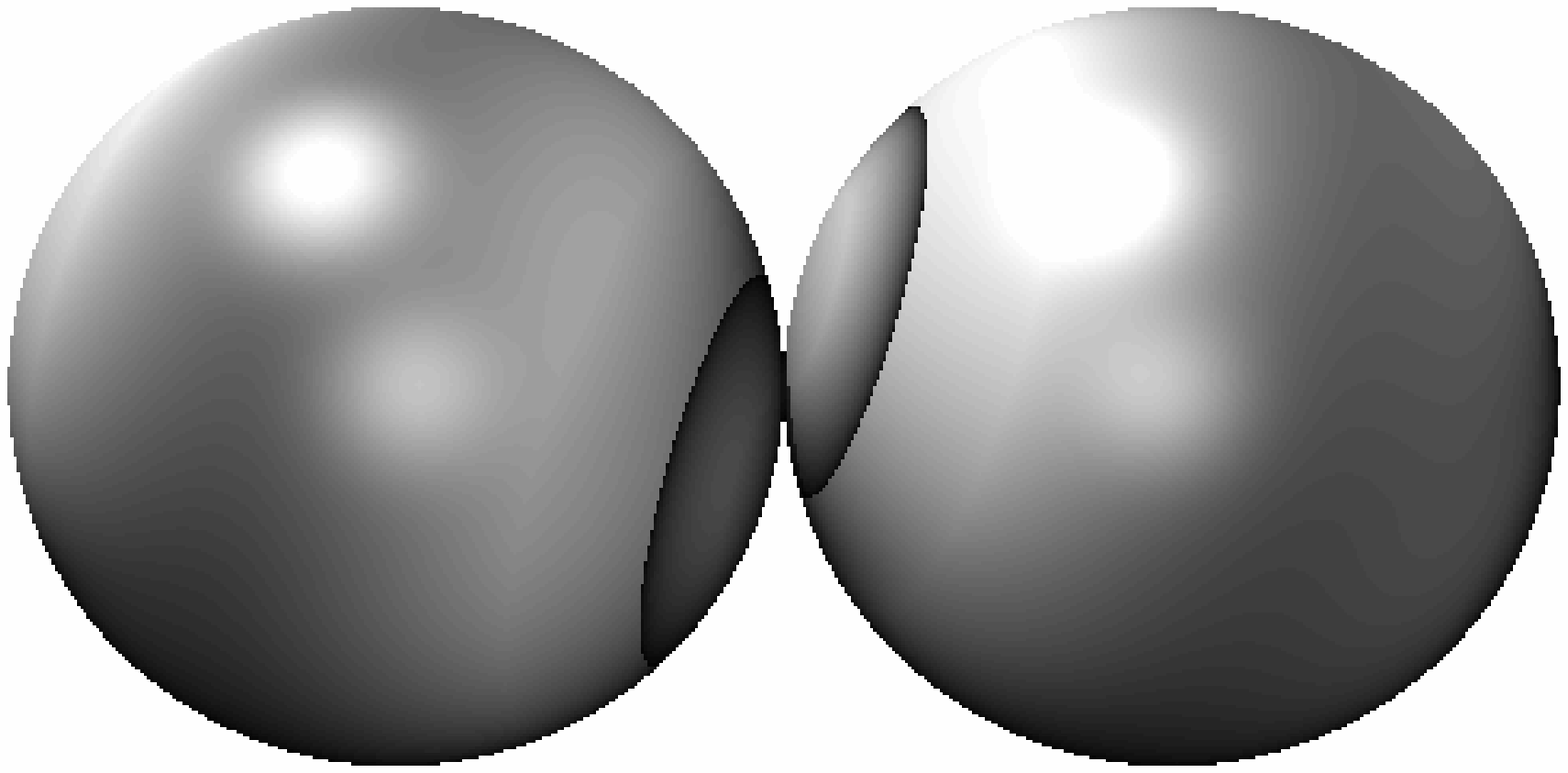}
\end{center}
\caption{}
\label{fig:spheres}
\end{figure}

In conjunction with Eq.~(\ref{SHS}), the adhesive part of the interaction
between two particles acts only 
if their point of contact lies inside a patch on each particle, as
depicted in Fig.~\ref{fig:spheres} for the case of a single patch ($n=1$).
Therefore, $\epsilon(1,2)\equiv \epsilon(\hat{\sss}_1,\hat{\sss}_2,\hat{\rrr}_{12})$
can be written as 
\bq
\epsilon(1,2)=\left\{\begin{array}{ll}
1 & \mbox{if}~ \hat{\sss}_1^{(p_1)}\cdot\hat{\rrr}_{12}\ge\cos\delta 
~\mbox{and}~ 
-\hat{\sss}_2^{(p_2)}\cdot\hat{\rrr}_{12}\ge\cos\delta~\mbox{for
some combination ($p_1,p_2$)} \\
0 & \mbox{otherwise.}
\end{array}
\right.
\label{epsilon}
\eq

Each patch occupies a portion of the sphere's surface 
covered by the solid angle $2\pi(1-\cos\delta)$ and a fundamental
role will be played in our discussion by the fraction of solid angle
(i.e., the coverage) associated with $\delta$, namely
\bq
\chi_0\left(\delta\right) = \frac{1}{2} \left(1-\cos\delta\right)=
\sin ^2 \left(\frac{\delta}{2}\right).
\label{coverage}
\label{chi}
\eq
%%%%%%%%%%%%%%%%%%%%%%%%%%%%%%%%%%%%%%%%%%%%%%%%%%%%%%%%%%%%%%%%%%%%%%%%%%%%%%%
\section{Analysis of the model}
\label{sect:analysis}
\subsection{Analytical solution}
We now tackle the analytical solution of this problem based on two simple
approximations: the virial expansion and the Cn class of closures. 
%%%%%%%%%%%%%%%%%%%%%%%%%%%%%%%%%%%%%%%%%%%%%%%%%%%%%%%%%%%%%%%%%%%%%%%%%%%%%%%
\subsubsection[]{Virial expansion}
%%%%%%%%%%%%%%%%%%%%%%%%%%%%%%%%%%%%%%%%%%%%%%%%%%%%%%%%%%%%%%%%%%%%%%%%%%%%%%%
As shown in Appendix B,
the first two virial coefficients for this model are
\bq
b_2&=&B_2/v_0=4-12\frac{\chi_1}{12\tau}~,\\
b_3&=&B_3/v_0^2=10-60\frac{\chi_1}{12\tau}+144\frac{\chi_2}{(12\tau)^2}-
96\frac{\chi_3}{(12\tau)^3}~,
\label{virial}
\eq
where $v_0=\pi\sigma^3/6$ is the volume of a sphere and
\bq \label{chi1}
\chi_1\left(\delta,n\right)&=& \left \langle \epsilon(2,3) 
\right \rangle_{\Omega_2,\Omega_3}\, \\ \label{chi2}
\chi_2\left(\delta,n\right)&=& 4~ \left \langle \epsilon(1,3) \epsilon(2,3) 
\Theta\left(\frac{\pi}{3}-\theta_{13}\right) \right \rangle_{\Omega_1,\Omega_2,\Omega_3,\Omega_{13}}
\, \\ \label{chi3}
\chi_3\left(\delta,n\right)&=&\left \langle \epsilon(1,2) \epsilon(1,3) \epsilon(2,3) 
\right \rangle_{\Omega_1,\Omega_2,\Omega_3}\vert_{\theta_{12}=\pi/3,\theta_{23}=2\pi/3} \,
\eq
where we have defined the angular average (with $d\tilde{\Omega}=d\Omega/4\pi$)
\bq
\left \langle \ldots \right \rangle_{\Omega} = \int ~d\tilde{\Omega} \ldots
\label{average}
\eq
Here, $\theta_{ij}$ is the angle between $\hat{\rrr}_{ij}$
and $\hat{\rrr}_{12}$ (which can be chosen along the $z$ axis), and 
$\epsilon(i,j)$ is always associated with a delta function that forces
spheres $i$ and $j$ to be in contact.
Note that in Eq.~(\ref{virial}) the effect of anisotropy is embedded in
the parameters $\chi_1,\chi_2,\chi_3$ defined in Eqs.~(\ref{chi1}),
(\ref{chi2}), and (\ref{chi3}), and that these parameters are therefore functions of $\delta$ and $n$.
The isotropic case is recovered when all $\chi$'s equal $1$.
We remark that the
expression for $\chi_2$ involves an average
over the relative orientations $\Omega_{13}$
while there is an overlap between
spheres 1 and 2, each of which is simultaneously in contact with sphere 3.
Under such conditions there is always a maximum possible angle $\pi/3$ for
$\theta_{13}$ and this gives rise to the normalization factor of $4$ in
Eq.~(\ref{chi2}).

If one limits the expansion to the second virial coefficient,
a law of corresponding states based on the rescaling $\tau\rightarrow
\tau/\chi_1$ between the patchy and the isotropic SHS
models holds true. This correspondence breaks down even at the level of the third 
virial. 

\begin{figure}[ht!]
\begin{center}
\includegraphics[width=8cm]{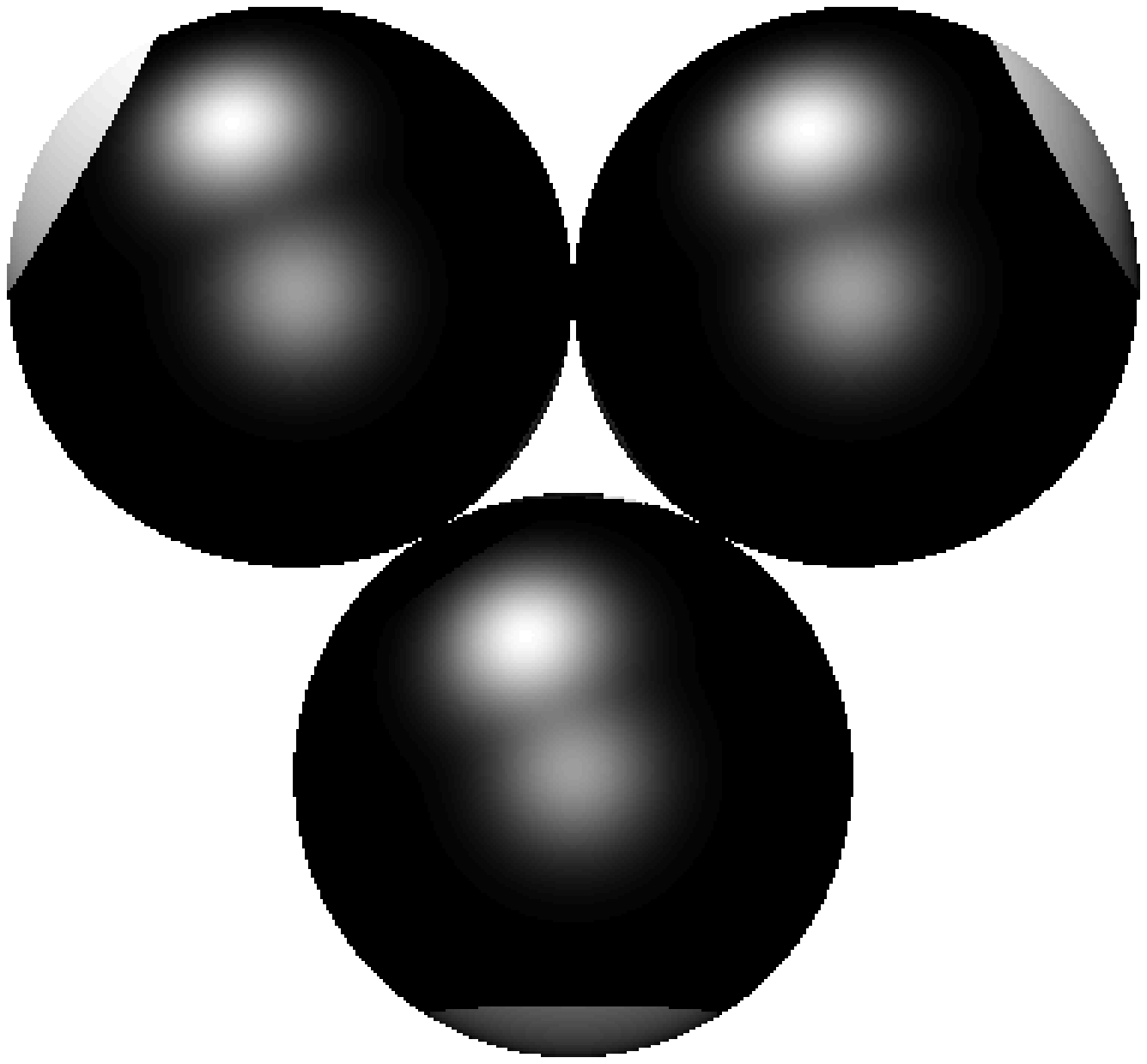}
\includegraphics[width=8cm]{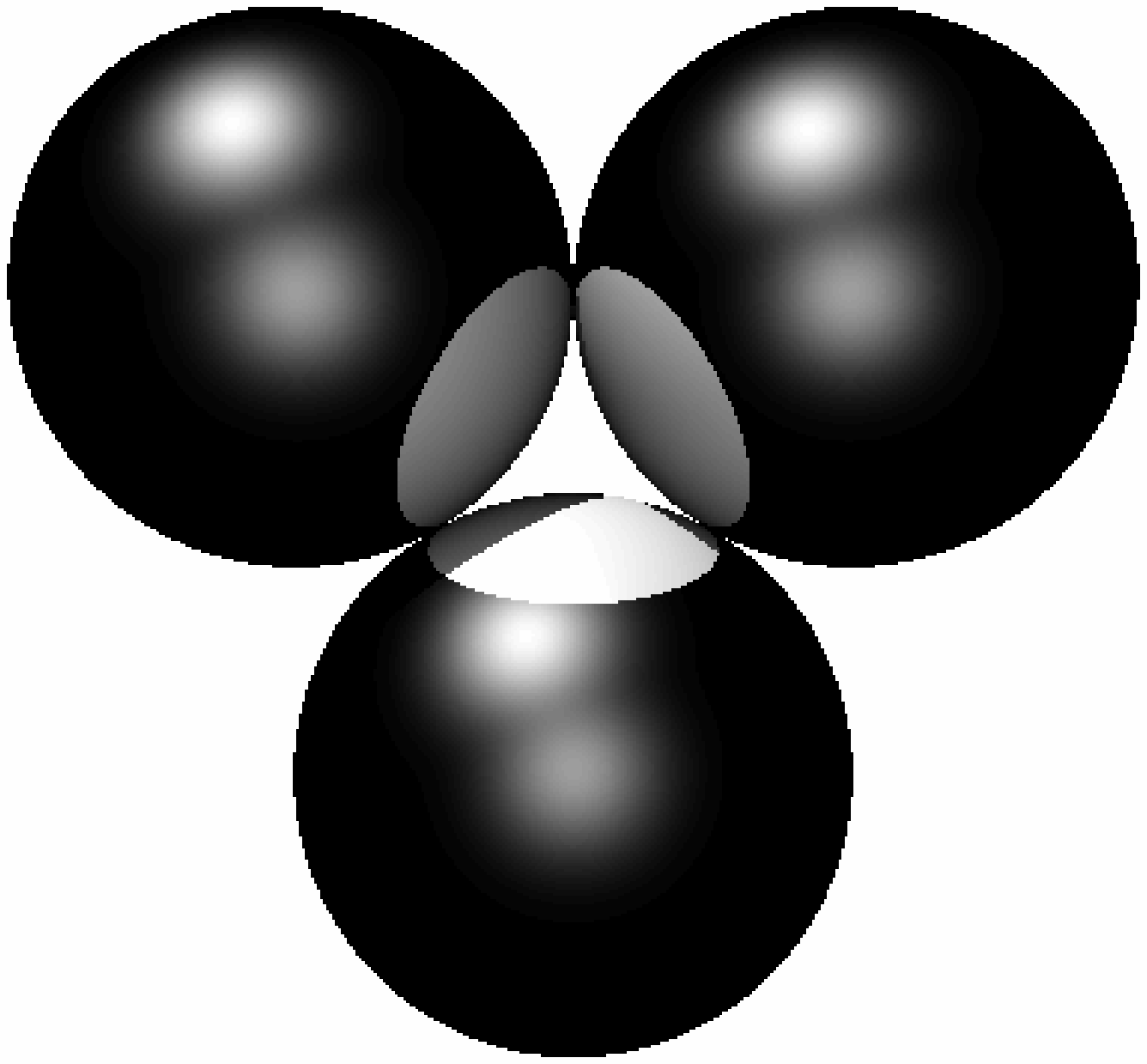}
\end{center}
\caption{}
\label{fig:5p6}
\end{figure}

\begin{figure}[ht!]
\begin{center}
\includegraphics[width=10cm]{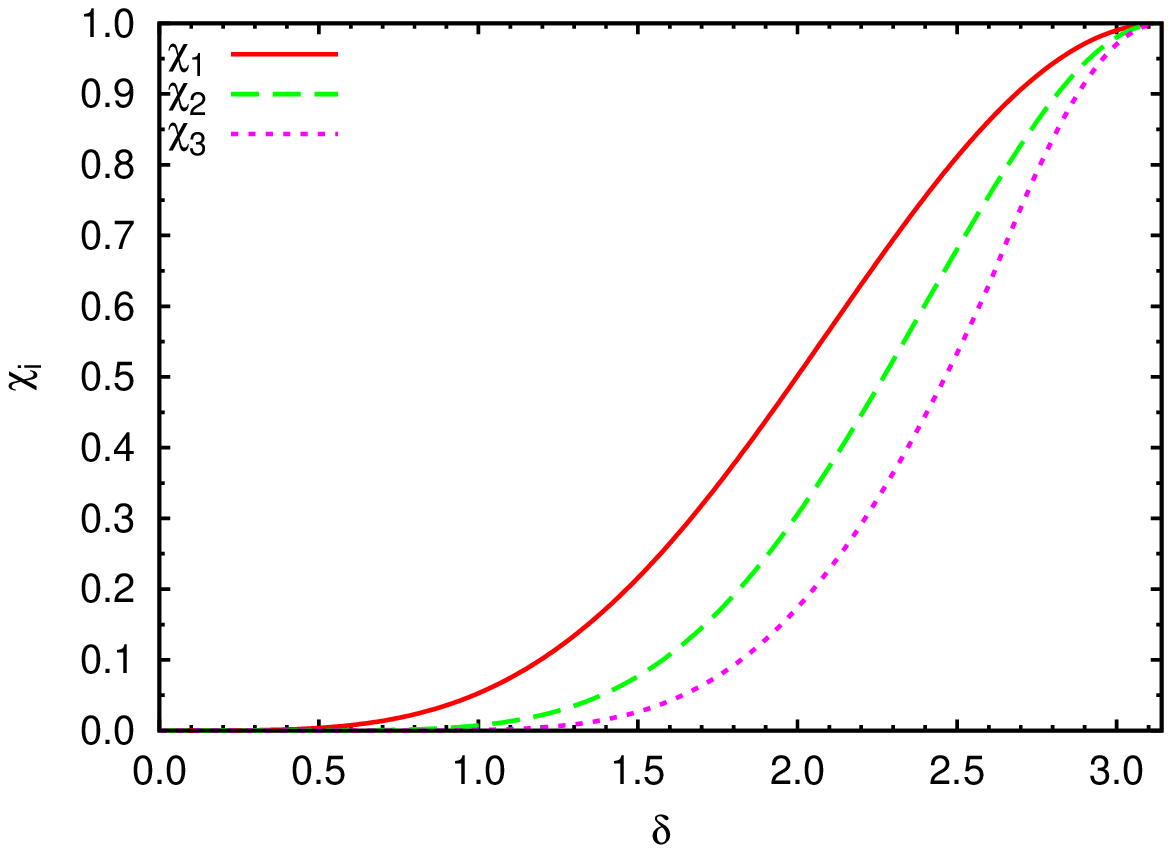}
\end{center}
\caption{}
\label{fig:chi1}
\end{figure}

It is easy to see that $\chi_1 = n^2 \chi_0^2$ as this is simply the product of the separate
coverages on each sphere. A calculation of $\chi_2$ and $\chi_3$ is much more laborious
and can be found in Appendix \ref{app:op} for the case of a single patch. The
final result in this case is
\bq \label{1c1}
\chi_1&=& \chi_0^2~,\\ \label{1c2}
\chi_2&=&\chi_0^2 Q_1(\delta)~,\\ \label{1c3}
\chi_3&=&R_1^3(\delta)~,
\eq
where the coefficients $Q_1$ and $R_1$ are given in Appendix
\ref{app:op}.
For $\delta\ge 5\pi/6$ it is possible to have three mutually bonded spheres with the
patch vectors pointing either inward and outward (see Fig.~\ref{fig:5p6}). 
Note that for the isotropic limit $\delta=\pi$ all $\chi_i$ $(i=1,2,3)$ are equal to $1$
as they should be. The three $\chi_i$ coefficients are plotted in
Fig. \ref{fig:chi1} as functions of $\delta$.

\begin{figure}[ht!]
\begin{center}
\includegraphics[width=8cm]{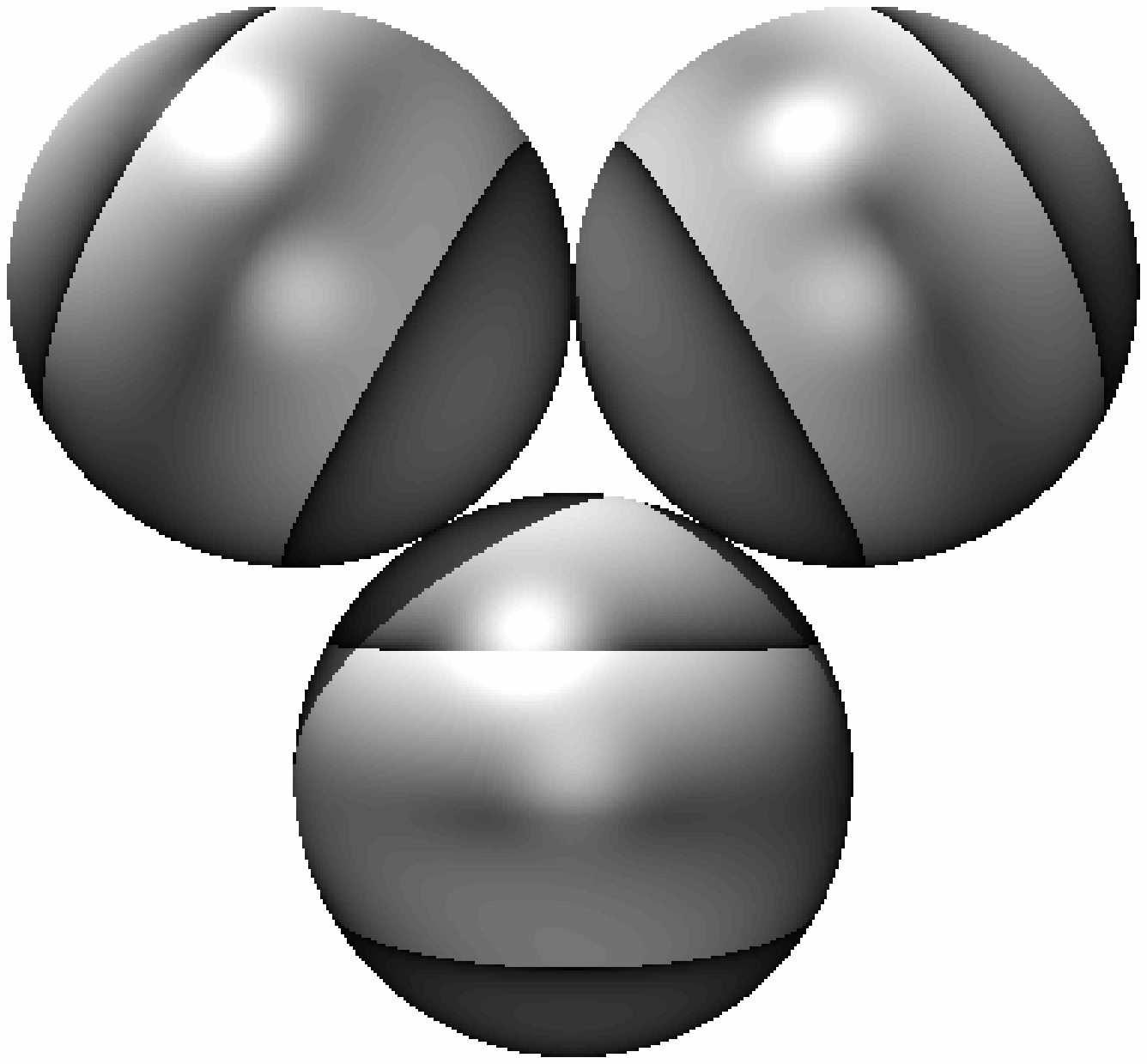}
\includegraphics[width=8cm]{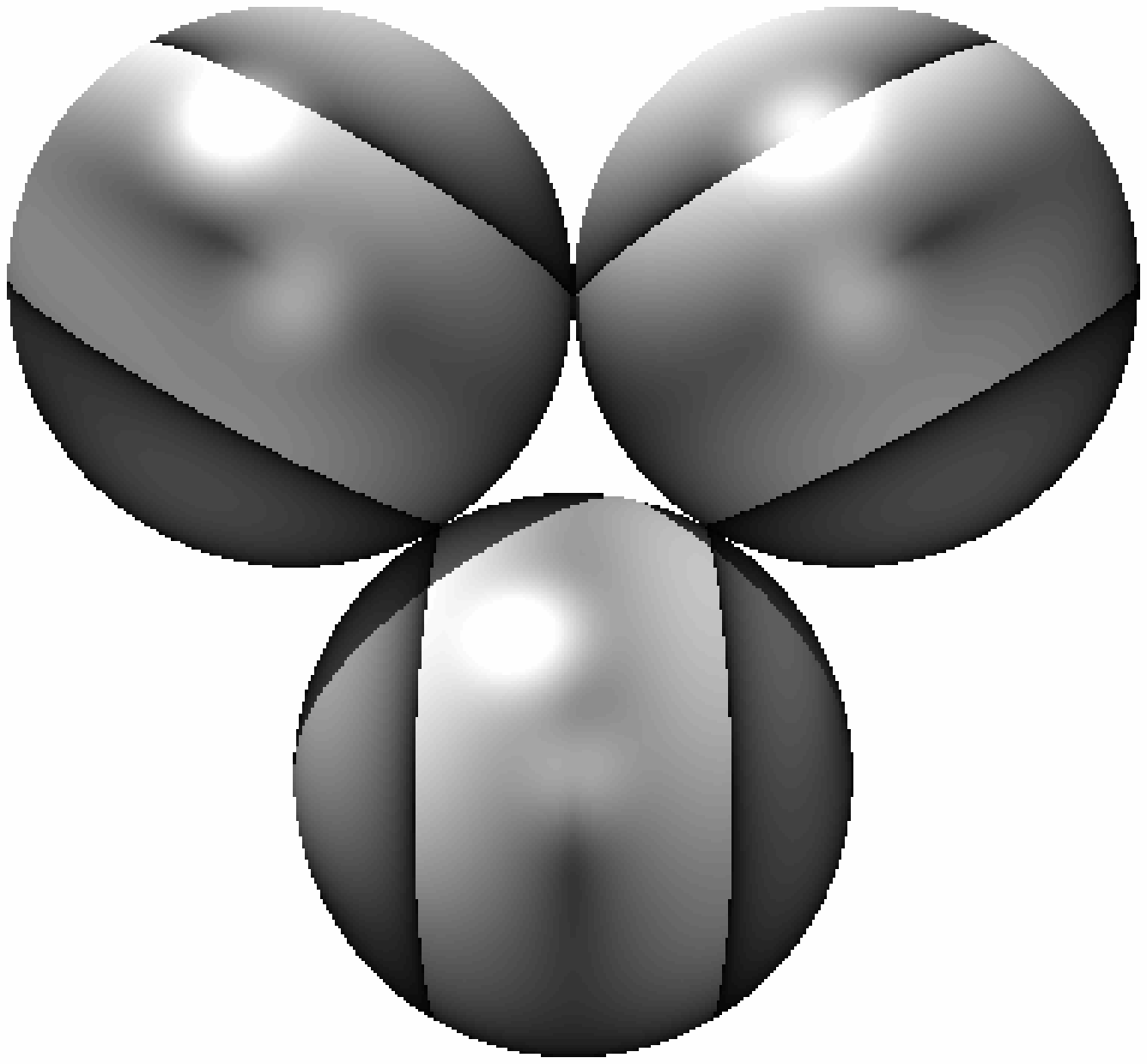}
\end{center}
\caption{}
\label{fig:3-fold}
\end{figure}

\begin{figure}[ht!]
\begin{center}
\includegraphics[width=10cm]{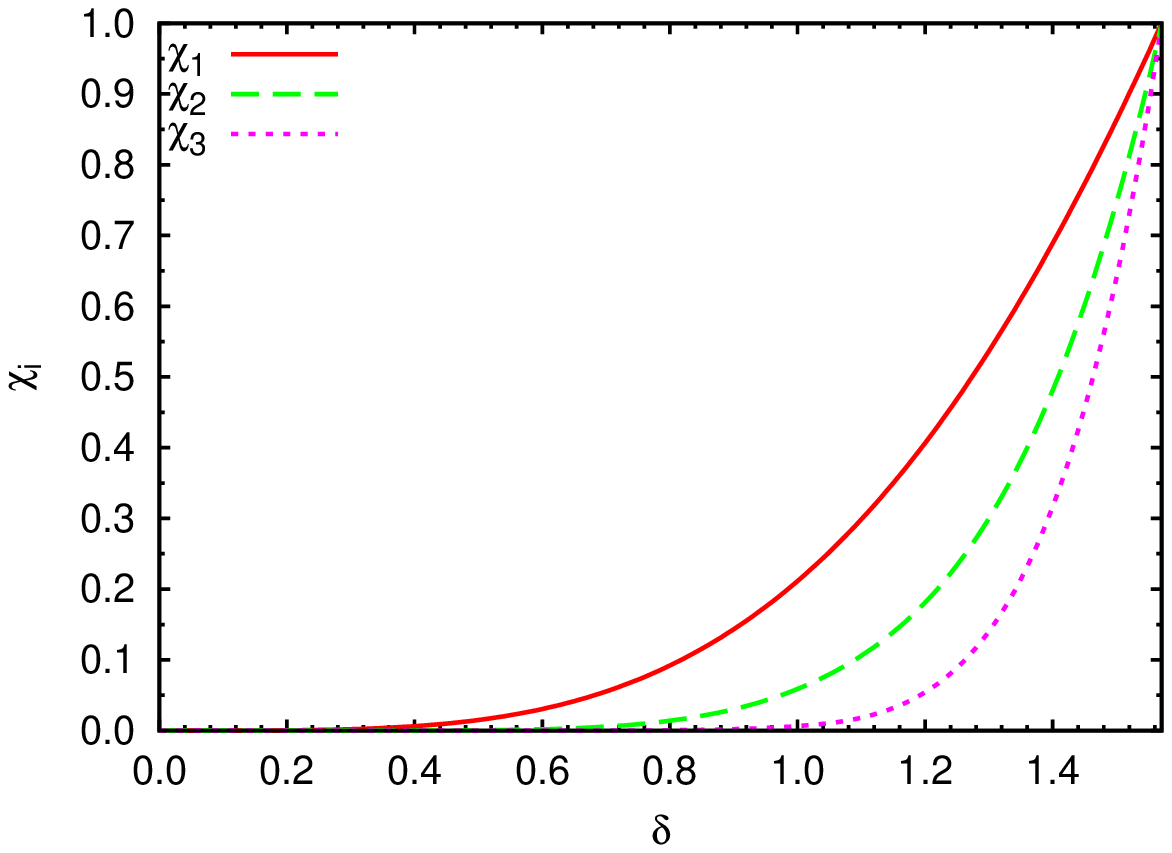}
\end{center}
\caption{}
\label{fig:chi2}
\end{figure}

For spheres with two diametrically opposite patches, each of width $\delta$, one finds
\bq \label{2c1}
\chi_1&=& 4 \chi_0^2~,\\ \label{2c2}
\chi_2&=& 4 \chi_0^2 Q_2(\delta)~,\\ \label{2c3}
\chi_3&=&R_2^3(\delta)~,
\eq
where the coefficients $Q_2$ and $R_2$ are given in Appendix
\ref{app:tp}.
Note that in this case when $\delta>\pi/3$ it is also possible to have sphere 1
in contact with spheres 2 and 3 through different patches as shown in
Fig.~\ref{fig:3-fold}. 
The three $\chi_i$ coefficients are plotted in
Fig.~\ref{fig:chi2} as functions of $\delta$.

The virial expansion of the excess free energy density is
\bq \label{f-virial}
\beta f^{ex}v_0=b_2\eta^2+\frac{1}{2}b_3\eta^3+\ldots~,
\eq
where $\eta=\rho v_0$ is the hard sphere packing fraction. This allows the calculation
of the corresponding pressure and chemical potential,
\bq \nonumber
\beta P(\tau,\eta) v_0 &=& \eta+b_2\eta^2+b_3\eta^3+\ldots,\\ \nonumber
\beta \mu(\tau,\eta) &=& \ln(\Lambda^3/v_0)+ \ln\eta+2b_2\eta+\frac{3}{2}b_3\eta^2+\ldots,
\eq
where $\Lambda$ is the de Broglie wavelength.
%%%%%%%%%%%%%%%%%%%%%%%%%%%%%%%%%%%%%%%%%%%%%%%%%%%%%%%%%%%%%%%%%%%%%%%%%%%%%%%
\subsubsection[]{Integral equations within the Cn closures}
%%%%%%%%%%%%%%%%%%%%%%%%%%%%%%%%%%%%%%%%%%%%%%%%%%%%%%%%%%%%%%%%%%%%%%%%%%%%%%%
While the virial expansion only allows a limited 
low-density region of the phase diagram to be probed, the integral equation
approach is much more powerful in this respect. The trade-off is, of course, that since
the OZ equation 
involves the total correlation function $h$ and direct correlation function $c$,
both of which are unknown,
it can be solved only after adding a closure, that is a second,
approximate, relationship involving $h$, $c$, and the pair potential.
In this section we discuss a particular class of these closures 
(denoted Cn hereafter) which have already been 
exploited in the isotropic case and have proven to provide reasonably good predictions 
even for intermediate densities.\cite{Gazzillo04}

The OZ equation for a homogeneous fluid of molecules
interacting through anisotropic pair potentials is
\bq
h(1,2)=c(1,2)+\rho\int d(3) c(1,3)h(3,2)\,~,
\label{OZ1}
\eq
where $d(i) \equiv d\rrr_id\tilde{\Omega}_i$.
More explicitly [see Eq.~(\ref{average})]
\bq
h(1,2)=c(1,2)+\rho\int d\rrr_3 \langle c(1,3)h(3,2)
\rangle_{\Omega_3}~.
\label{OZ2}
\eq

In a homogeneous fluid, translational invariance of any
correlation function implies that one can introduce reduced coordinates $\rrr_{12}=
\rrr_2-\rrr_1$ and $\rrr_{13} = \rrr_{3}-\rrr_{1}$
\bq
h(\rrr_{12},\Omega_1,\Omega_2)=c(\rrr_{12},\Omega_1,\Omega_2)+
\rho\int d\rrr_3 \langle c(\rrr_{13},\Omega_1,\Omega_3)
h(\rrr_{32},\Omega_3,\Omega_2)
\rangle_{\Omega_3}~.
\label{OZ3}
\eq
The presence of the convolution makes it convenient to Fourier
transform this equation with respect to the position
variable $\rrr$. This yields for the corresponding functions (indicated
with a hat) in Fourier space $\kkk$
\bq
\hat{h}(\kkk,\Omega_1,\Omega_2)=\hat{c}(\kkk,\Omega_1,\Omega_2)
+\rho\langle\hat{c}(\kkk,\Omega_1,\Omega_3)\hat{h}(\kkk,\Omega_3,\Omega_2)
\rangle_{\Omega_3}~.
\label{OZ4}
\eq
The additional complication with respect to the isotropic case is the presence
of the orientational average of the product appearing in Eq.~(\ref{OZ4}). In order
to make progress, we use a simple angular decoupling approximation  
\bq
\langle\hat{c}(\kkk,\Omega_1,\Omega_3)\hat{h}(\kkk,\Omega_3,\Omega_2)
\rangle_{\Omega_3}\simeq
\langle\hat{c}(\kkk,\Omega_1,\Omega_3)\rangle_{\Omega_3}
\langle\hat{h}(\kkk,\Omega_3,\Omega_2)\rangle_{\Omega_3}~.
\label{OMF}
\eq
As discussed in Ref.~\onlinecite{Gazzillo04}, Cn closures are based on a density
expansion of the cavity function $y(1,2)$ which is related to the radial distribution
function $g(1,2)$ by $g(1,2)=y(1,2) e(1,2)$. 

In the presence of anisotropy, all correlation function clearly depend upon
the solid angles $\Omega_1,\Omega_2,\Omega_{12}$. It is then customary to
consider\cite{Gray} the corresponding angular averaged quantities 
$g(r\equiv r_{12})= \langle g(1,2) \rangle_{\Omega_1,\Omega_2,\Omega_{12}}$
and similarly for $y(r \equiv r_{12})$. Within Cn closures, for
$r>\sigma$, the radial
distribution function $g(r)$ coincides with the cavity function $y(r)$. 
A density expansion of the cavity function yields 
\bq
y(r)=1+\rho y_1(r)+\ldots~,
\label{expansion}
\eq
where
\bq \label{y1}
y_1(r_{12})=\int  d\rrr_3 \left \langle f(1,3)f(3,2) 
\right \rangle_{\Omega_1,\Omega_2,\Omega_3,\Omega_{12}}
\eq
Calculation of Eq.~(\ref{y1}) proceeds using arguments akin to those presented in 
Appendix \ref{app:ve}, which are based on the decomposition in Eq.~(\ref{mayer}).
The integral in Eq.~(\ref{y1})
then splits into three integrals containing the various combinations of
the HS and the sticky parts of the Mayer function as in Eq.~(\ref{mayer}). 
%%%%%%%%%%%%%%%%%%%%%%%%%%%%%%%%%%%%%%%%%%%%%%%%%%%%%%%%%%%%%%%
\subsection{Monte Carlo algorithms for Baxter-like potentials}
%%%%%%%%%%%%%%%%%%%%%%%%%%%%%%%%%%%%%%%%%%%%%%%%%%%%%%%%%%%%%%%
Monte Carlo simulations of adhesive hard spheres require particular
care even in the isotropic case because of the singular nature of the potential.
For completeness we summarize the main ideas below, deferring to Ref.~\onlinecite{Miller07}
for the details.

Conventional Monte Carlo
displacements of a SHS would fail because the bonded states between particles
occupy an infinitesimal volume of configuration space (and so would never be located
by random displacements) but have infinite strength (and so would never be broken).
The solution is to compare the integrated weights of the various bonded and unbonded
states, which are finite.  Specialized algorithms that exploit this approach have been
devised for the canonical ensemble\cite{Seaton87,Kranendonk88} and were subsequently
extended to the grand canonical ensemble.\cite{Jamnik94,Miller04}  The latter is
particularly convenient for identifying the critical point.\cite{Miller03}
\par
The Monte Carlo algorithm for isotropic adhesive spheres can be modified to deal with
the patchy case by incorporating the anisotropy in the acceptance criterion for trial
moves.  Trial moves are attempted as described in detail in Ref.~\onlinecite{Miller04} as though
the spheres were uniformly adhesive.  Once the trial position of the displaced
particle has been chosen, a uniformly distributed random orientation is selected.
The move is then accepted only if an overlap of hard cores is not generated (as in
the isotropic case) {\em and} if all contacts specified in the trial configuration have
patches suitably aligned to make the required bonds.  This scheme produces the desired
Boltzmann distribution\cite{Miller07} and is applicable to an arbitrary arrangement of
patches.  However, it becomes inefficient when the total adhesive coverage of the sphere
is small because the random generation of orientations is then unlikely to lead to
patches being aligned with bonds, leading to a high rejection rate.

%%%%%%%%%%%%%%%%%%%%%%%%%%%%%%%%%%%%%%%%%%%%%%%%%%%%%%%%%%%%%%%%%%%%%%%%%%%%%%%
\section{Structure}
\label{sect:structure}
%%%%%%%%%%%%%%%%%%%%%%%%%%%%%%%%%%%%%%%%%%%%%%%%%%%%%%%%%%%%%%%%%%%%%%%%%%%%%%%
In the following we shall compare predictions
from the combined C1-orientational mean field approximation and virial expansion with
the results of Monte Carlo simulations.

One finds that $y_1(r)$ is 
different from zero only in the region $0\le r\le 2 \sigma$ and 
\bq
y_1(r)=\frac{\pi}{12}\left(\frac{r}{\sigma}+4\right)\left(\frac{r}{\sigma}-2\right)^2+
\frac{\chi_1}{12\tau}2\pi\left(\frac{r}{\sigma}-2\right)
+\frac{\bar{\bar{\chi}}_2(r)}{(12\tau)^2}2\pi \frac{\sigma}{r},
\eq
where
\bq
\bar{\bar{\chi}}_2(r)&=& \left \langle \epsilon(1,2)\epsilon(1,3) 
\right\rangle_{\Omega_1,\Omega_2,\Omega_3} \vert_{\theta_{12}=2\arcsin(r/2\sigma)}.
\label{chibar}
\eq

In order to compute the angularly averaged radial distribution function $g(r)$ we have
solved the full OZ equation (\ref{OZ3}) within the C1 closure supplemented with
the decoupling approximation (\ref{OMF}).  The Wertheim-Baxter method\cite{Baxter68}
and the C1 closure combine such that only the cavity function at contact
depends upon the angular coefficients $\chi_1,\chi_2,\chi_3$. The solution for Baxter
function is 
\bq
q(r)&=&[a(r^2-\sigma^2)/2+b\sigma(r-\sigma)+q_\sigma\sigma^2]
\Theta(\sigma-r)~~~r>0
\eq
where
\bq
a&=&\frac{1+2\eta}{1-\eta}^2-\frac{12q\eta}{1-\eta}~,\\
b&=&-\frac{3\eta}{2(1-\eta)^2}+\frac{6q\eta}{1-\eta}~,\\
q_\sigma&=&\frac{\bar{y}^{C1}}{12\tau}~,\\
\bar{y}^{C1}&=&\langle
y^{C1}\left(r_{12}=\sigma,\Omega_1,\Omega_2,\Omega_{12}\right)\epsilon(1,2)
\rangle_{\Omega_1,\Omega_2,\Omega_{12}}=y_0+y_1\eta~.
\eq
The coefficients $y_{0,1}=y_{0,1}(\tau)$ are related to the reduced virial coefficients $b_{2,3}$
by Eq.~(\ref{yn_bn}), below.
Therefore we can read off their values
\bq
y_0&=&\chi_1~,\\
y_1&=&\left(30\frac{\chi_1}{12\tau}-144\frac{\chi_2}{(12\tau)^2}+
144\frac{\chi_3}{(12\tau)^3}\right)\tau~.
\label{y01}
\eq

In order to extract the numerical behavior of the radial
distribution function $g(r)$ we have employed
a discretization method due to Perram\cite{Perram75} to compute the numerical integral 

\bq
rh(r)=-q'(r)+2\pi\rho\int_0^\infty du\, q(u)(r-u)h(|r-u|)~.
\eq

\begin{figure}[ht!]
\begin{center}
\includegraphics[width=10cm]{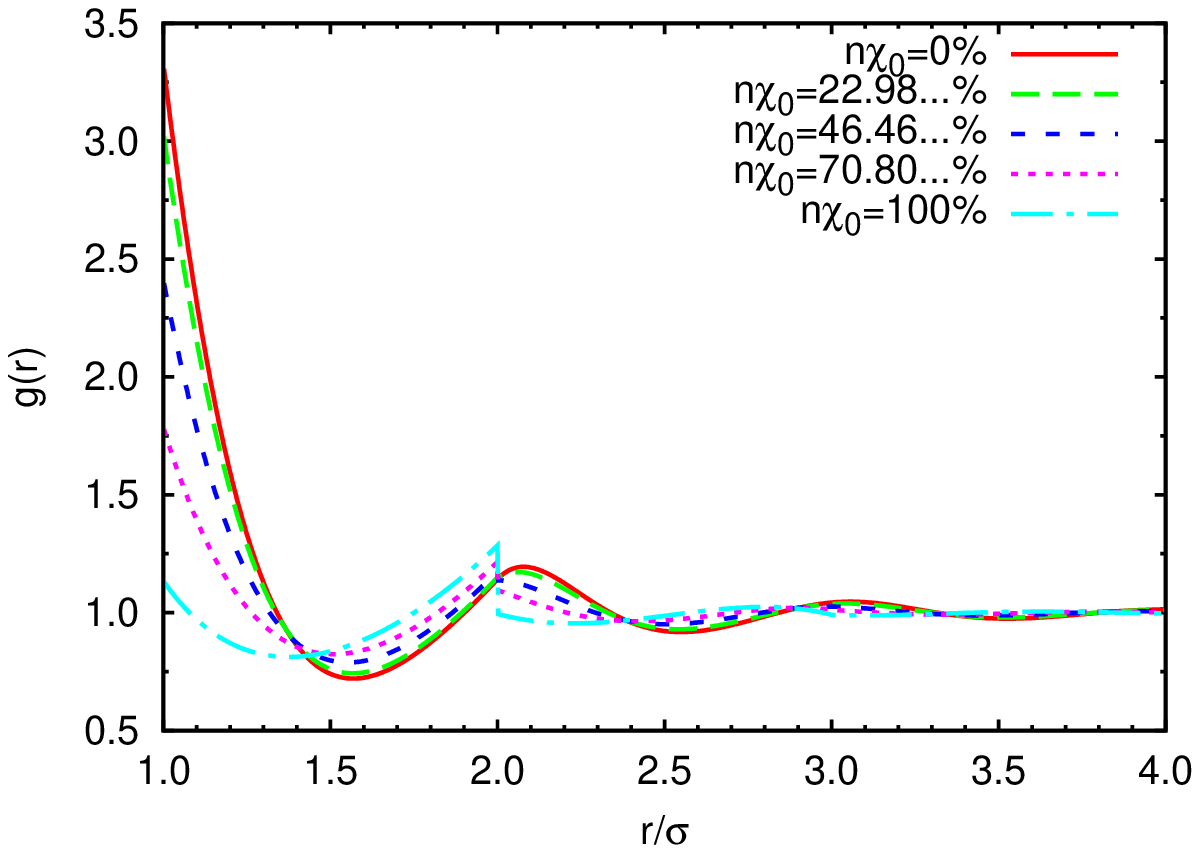}
\end{center}
\caption{}
\label{fig:gr1}
\end{figure}

\begin{figure}[ht!]
\begin{center}
\includegraphics[width=8cm]{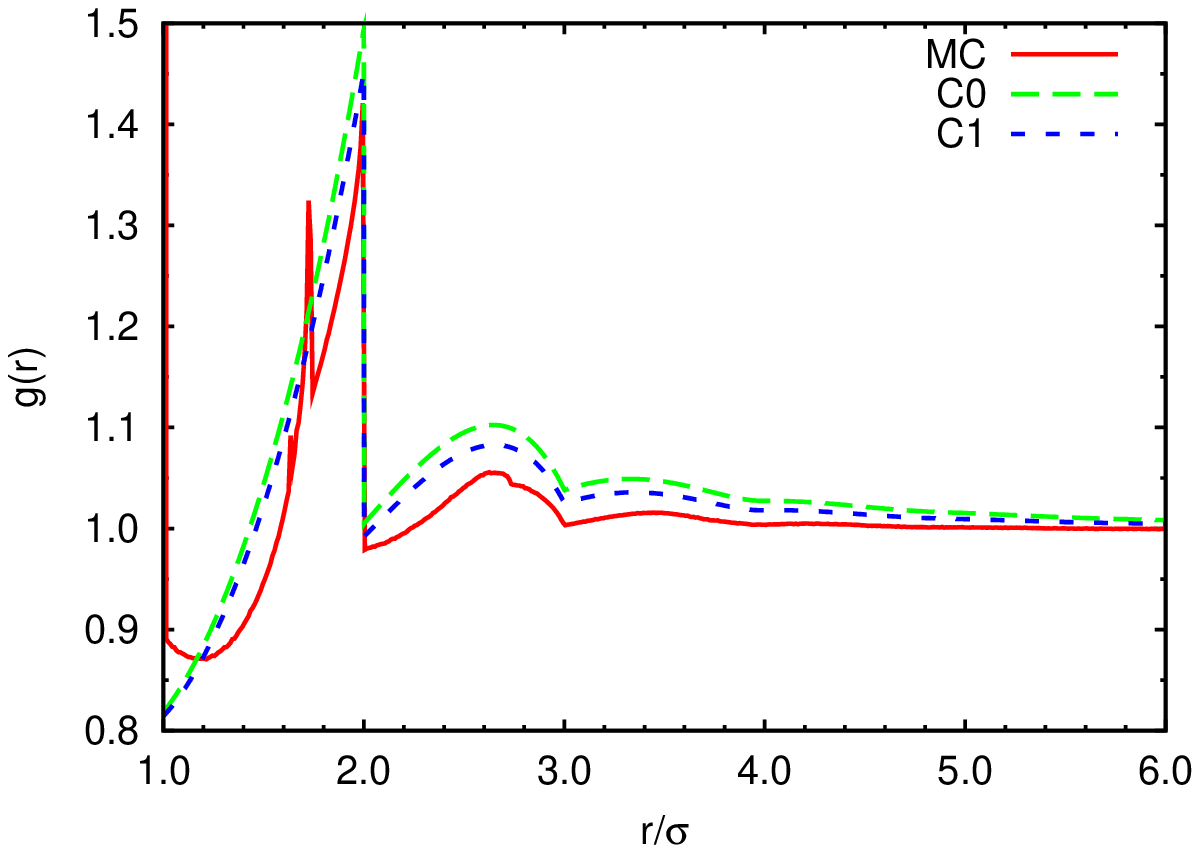}
\includegraphics[width=8cm]{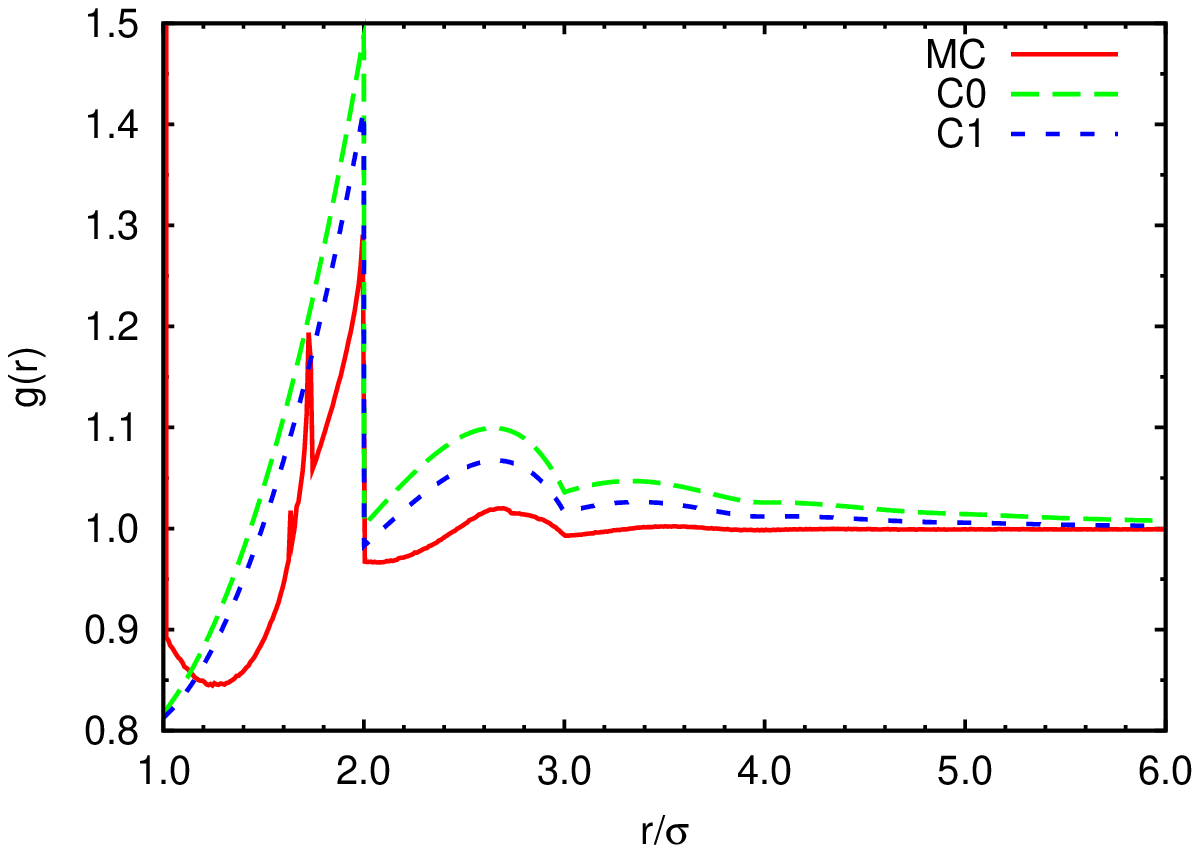}
\end{center}
\caption{}
\label{fig:grmc}
\end{figure}

For the one-patch spheres the result is reported in
Fig.~\ref{fig:gr1} for various values of $\delta$, at $\eta=0.4$ and
$\tau=0.2$. The choices $\delta=0$ and $\delta =\pi$ correspond to the 
limiting cases of pure HS and isotropic SHS, respectively. Upon decreasing the
size of the patch, the behavior smoothly interpolates between these two cases, as 
expected. The characteristic jump in $g(r)$ at $r=2 \sigma$ in the isotropic SHS
model\cite{Miller04} can be explicitly computed within the C1 integral equation closure
to be
\bq
g(2\sigma^+)-g(2\sigma ^-)=-6\eta [\bar{y}^{C1}/(12\tau)]^2~.
\label{jump}
\eq
The jump is also present for
intermediate values 
of $\delta$ and gradually fades out toward the HS result as illustrated in 
Fig.~\ref{fig:gr1}.
In order to assess the precision of the predictions of the Cn closures,
in Fig.~\ref{fig:grmc} we compare the radial distribution generated by
both C0 and C1 closures with Monte Carlo simulations (right-hand panel) and
with the corresponding isotropic case (left-hand panel).
The C1 closure is seen to follow the Monte Carlo behavior well over the
range of the ratio $r/\sigma$ considered, in both the isotropic and anisotropic
cases. 

Like the Percus-Yevick solution of the isotropic model, the C0 and C1
closures fail to capture certain $\delta$-function and step discontinuities 
in the radial distribution function,\cite{Cummings76} such as those visible 
in the Monte Carlo results in the range $\sigma<r<2\sigma$ in Fig. \ref{fig:grmc}.  
These features arise from clusters in which the distance between two 
particles is fixed or limited indirectly by a sequence of adhesive bonds, 
for example, the outermost pair of particles in face-sharing tetrahedra.  
These clusters are sampled correctly by the Monte Carlo simulations.\cite{Miller04_b}

%%%%%%%%%%%%%%%%%%%%%%%%%%%%%%%%%%%%%%%%%%%%%%%%%%%%%%%%%%%%%%%%%%%%%%%%%%%%%%%
\section{Fluid-Fluid coexistence curve}
\label{sect:coexistence}
%%%%%%%%%%%%%%%%%%%%%%%%%%%%%%%%%%%%%%%%%%%%%%%%%%%%%%%%%%%%%%%%%%%%%%%%%%%%%%%
An interesting issue,
both from the theoretical view point and for the
possible implications in predictions of experimental phase transitions
in solutions of globular proteins,
is the determination of the fluid-fluid coexistence
curve for the patchy sticky hard spheres, which we now address.  We note
that the dense fluid (liquid) phase, though often long-lived, is generally only
metastable for systems of particles interacting through sufficiently short-ranged
isotropic attractive forces.  However, it has recently been predicted that a
thermodynamically stable liquid will be recovered if the coordination
number of the particles is restricted to a maximum of six or less.\cite{Charbonneau07}
Nevertheless, it seems that the specific details of the
interactions must be taken into account before a firm conclusion can be drawn
for a particular model.\cite{Romano07}
%%%%%%%%%%%%%%%%%%%%%%%%%%%%%%%%%%%%%%%%%%%%%%%%%%%%%%%%%%%%%%%%%%%%%%%%%%%%%%%
\subsection{Virial expansion} 
%%%%%%%%%%%%%%%%%%%%%%%%%%%%%%%%%%%%%%%%%%%%%%%%%%%%%%%%%%%%%%%%%%%%%%%%%%%%%%

In order to find the coexistence or binodal line we need 
to solve for $\eta_1(\tau)$ and $\eta_2(\tau)$ the following set of 
equations,
\bq \nonumber 
P(\tau,\eta_1)&=&P(\tau,\eta_2)~,\\ \nonumber
\mu(\tau,\eta_1)&=&\mu(\tau,\eta_2)~.
\eq

A straightforward use of Eq.~(\ref{f-virial}) for the excess free 
energy density to this aim, however, yields meaningless results even at moderate
densities as one could have expected from the outset.
A way out of this problem was proposed in Ref.~\onlinecite{Fantoni06b}
in the context of polydisperse SHS fluids.
The idea hinges on a modification of the Carnahan-Starling expression for the
HS excess free energy  density,\cite{Carnahan69}
\bq
\beta f^{ex}_{cs}v_0=\frac{4-3\eta}{(1-\eta)^2}\eta^2~,
\label{f-cs}
\eq
so that it matches the patchy SHS result up to third order in
density. A possible choice is 
\bq \label{f-csm}
\beta f^{ex}v_0=(c-1)\eta\ln(1-\eta)+3d\frac{\eta^2}{1-\eta}
+c\frac{\eta^2}{(1-\eta)^2}~,
\eq
where $c$ and $d$ are parameters to be determined by
expanding to third order in density and matching to Eq.~(\ref{f-virial}).
We then find
\bq
c&=&\frac{b_3-2b_2+1}{3}~,\\
d&=&\frac{b_2-1}{3}~.
\eq
The 
pressure and the chemical potential are then
\bq \nonumber
\beta P v_0 &=&
\eta+\eta^2\frac{1+3d(1-\eta)+\eta[\eta-2+c(3-\eta)]}{(1-\eta)^3}~,\\ 
\nonumber
\beta \mu &=& \ln(\Lambda^3/v_0)+\ln\eta+(c-1)\ln(1-\eta)+\frac{(1+c+6d)\eta-(2-2c+9d)\eta^2
+(1-c+3d)\eta^3}{(1-\eta)^3}~,
\eq
respectively. In the limit $c=d=1$ Eq.(\ref{f-csm}) reduces to Eq.(\ref{f-cs}) as expected.

\begin{figure}[ht!]
\begin{center}
\includegraphics[width=8cm]{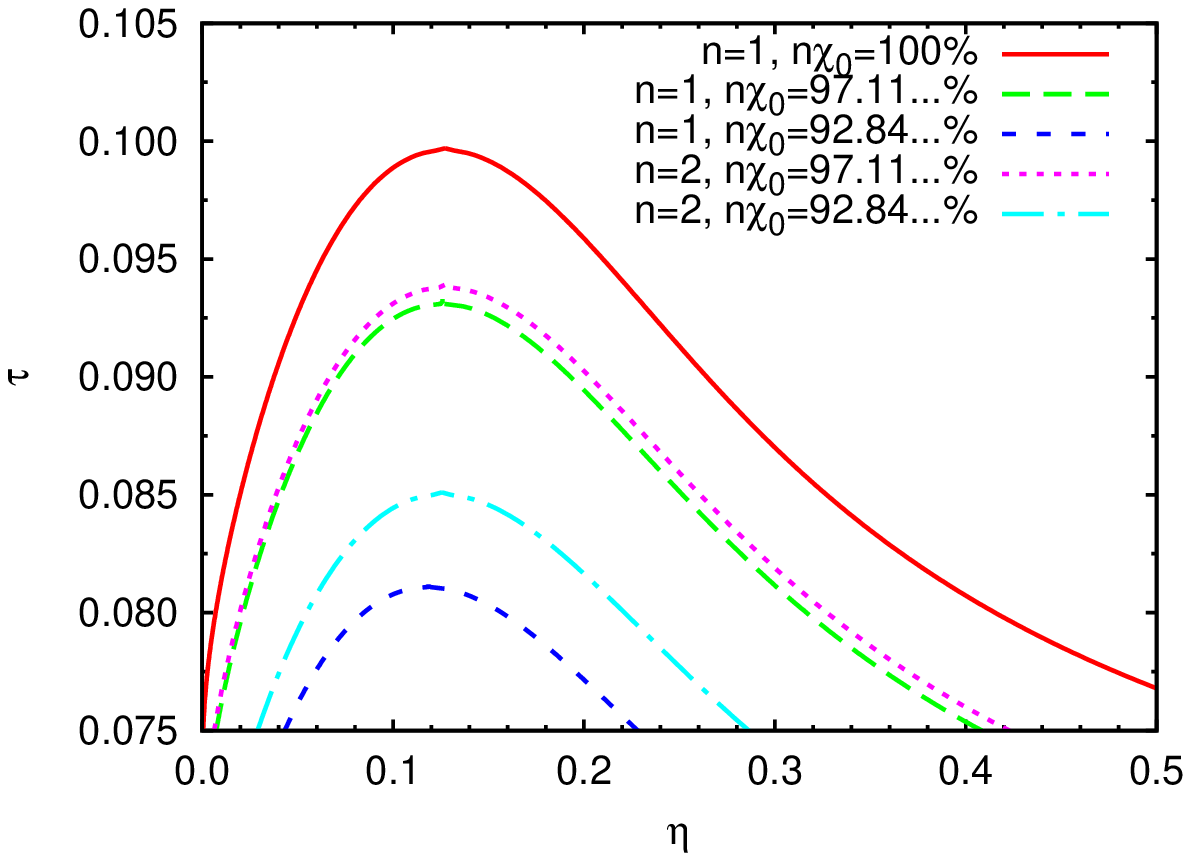}
\includegraphics[width=8cm]{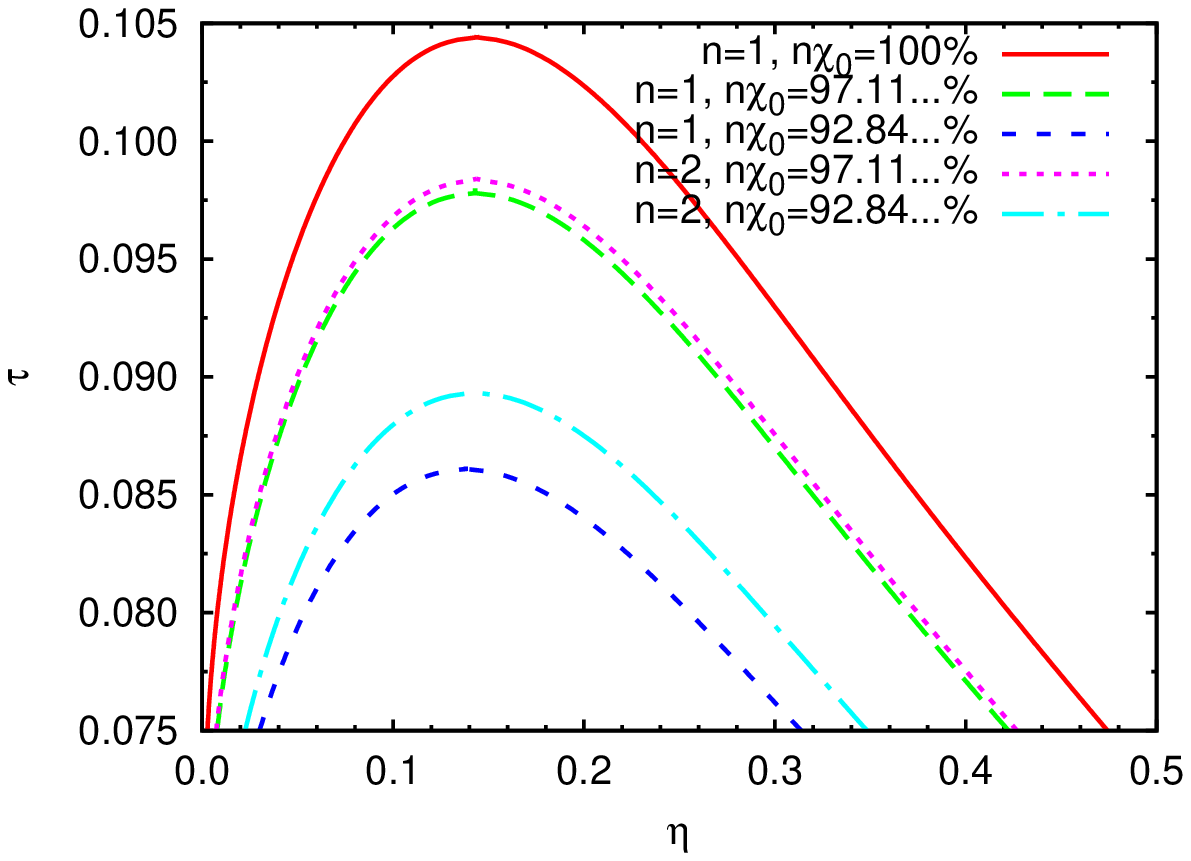}
\end{center}
\caption{}
\label{fig:binodal1c1}
\end{figure}

The behavior of the binodal line as a function of $\delta$ is shown in 
Fig. \ref{fig:binodal1c1}. As $\delta$ decreases, the coexistence region
shrinks as expected, since HS fluids ($\delta=0$) admit only a single
phase.\cite{Alder57}

%%%%%%%%%%%%%%%%%%%%%%%%%%%%%%%%%%%%%%%%%%%%%%%%%%%%%%%%%%%%%%%%%%%%%%%%%%%%%%%
\subsection[]{C1 integral equation}
%%%%%%%%%%%%%%%%%%%%%%%%%%%%%%%%%%%%%%%%%%%%%%%%%%%%%%%%%%%%%%%%%%%%%%%%%%%%%%%
An alternative route is to start form the excess free energy stemming from
the energy route of the C1 approximation\cite{Gazzillo04}
\bq \label{f-c1}
\beta f^{ex}v_0=\beta f_{cs}^{ex}v_0-(b_2-b_2^{\rm HS})\eta^2
+(b_3-b_3^{\rm HS})\frac{\eta^3}{2}~,
\eq
where $b_n^{\rm HS}=b_n(\tau\rightarrow\infty)$. The rescaled virial coefficients
$b_n=b_n(\tau)$ can be related to the values of the corresponding
coefficients of the expansion for the cavity function at contact
$\bar{y}=y_0+y_1\eta+y_2\eta^2+\ldots$  
by means of the relation (see e.g., Ref.~\onlinecite{Gazzillo04} and references therein):
\bq
y_{n-2}(\tau)/\tau^2=
\frac{1}{n-1}\frac{d[b_n(\tau)-b^{\rm HS}_n]}{d\tau},~~~n\ge 2~.
\label{yn_bn}
\eq
Hence, we have
\bq
\beta Pv_0&=&\eta+2\eta^2\frac{\eta-2}{(\eta-1)^3}
-\frac{\chi_1}{\tau}\eta^2
+\left(-5\frac{\chi_1}{\tau}+\frac{\chi_2}{\tau^2}-
\frac{1}{18}\frac{\chi_3}{\tau^3}\right)\eta^3~,\\
\beta\mu&=&\ln(\Lambda^3/v_0)+\ln\eta+\eta\frac{8+3(\eta-3)\eta}{(1-\eta)^3}
-\frac{\chi_1}{\tau}2\eta
+\left(-5\frac{\chi_1}{\tau}+\frac{\chi_2}{\tau^2}-
\frac{1}{18}\frac{\chi_3}{\tau^3}\right)\frac{3\eta^2}{2}~.
\eq
We remark that these results need no further orientational 
approximations as all effects of anisotropy are exactly included
in $\chi_{1,2,3}$. 

\begin{figure}[ht!]
\begin{center}
\includegraphics[width=10cm]{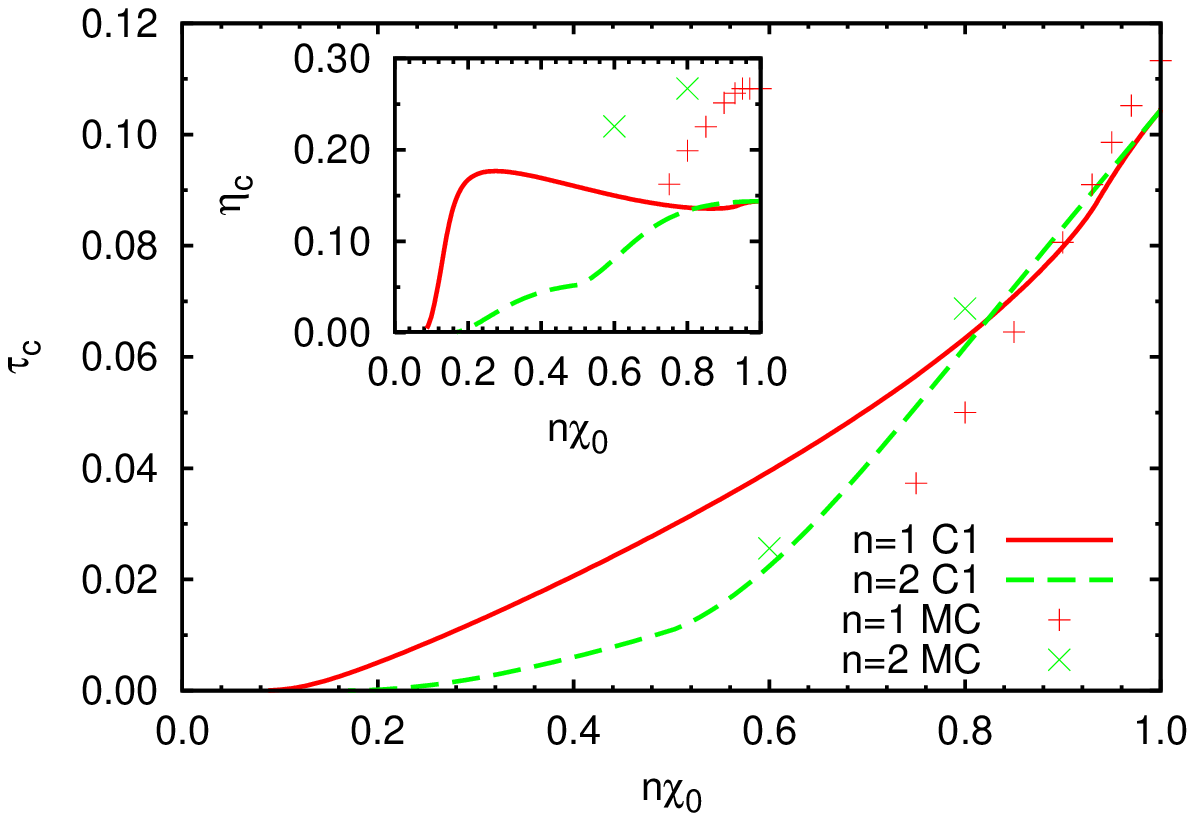}
\end{center}
\caption{} 
\label{fig:critical1c1}
\end{figure}

The results for the binodals are shown in Fig. \ref{fig:binodal1c1}.
For each of the adhesive coverages depicted, both theoretical treatments
predict that
the binodal line for the two-patch model lies above its counterpart
for a single patch. This could be expected
on physical grounds, since a more distributed region of adhesion usually facilitates
the aggregation process. A closer analysis, however, indicates that this is
not always the case. 
This is shown in Fig.~\ref{fig:critical1c1} where we report the change in the critical
point as a function of the adhesive coverage $n\chi_0$ of the sphere surface. 
The difference between one and two patches decreases as $\delta$ increases, 
as expected, but it is clearly visible through the whole range of existence.
Remarkably, there is an inversion of the two curves around $80\%$ coverage. For 
approximately $n\chi_0 \ge 0.8$ the critical temperature for two patches lies
above its one-patch counterpart. This means that the fluid-fluid transition line
is encountered at higher temperature when an identical adhesive coverage is distributed
over two spots rather than consolidated into a single big patch. However, this is
no longer true when the size of the patch becomes too small. The reason
is that under such conditions, it is then possible to bind three or more particles
within a single patch, whereas at most two particles (one on each of the two opposite patches)
can be attached in the two-patch case.

An additional noteworthy feature of Fig.~\ref{fig:critical1c1} is the
existence of a kink at $\delta=\pi/3$ ($50\%$ coverage)
in the two-patch case. The kink is related to the degeneracy
illustrated in Fig.~\ref{fig:3-fold}. A similar kink occurs in the
single-patch curve at $\delta=5 \pi/6$ ($93.3\%$ coverage). Again,
this stems from a degeneracy, as illustrated in Fig.~\ref{fig:5p6}.

Due to the inherent difficulty of tracing out critical temperatures
to low coverage, we have been unable to verify the crossing of the one- and
two-patch critical temperatures
by Monte Carlo simulations.  Fig.~\ref{fig:critical1c1} reports MC results
down to around $60\%$ coverage and the two-patch critical temperature is always above
the one-patch case. We suspect, however, that an inversion
might still occur in MC simulations, but at lower values of $\chi_0$, as yet inaccessible to our simulations.

The sensitivity of the shape and location of the coexistence curve to the
geometry of the adhesive distribution is quite a remarkable feature of this archetypal
patchy model.  It seems likely, therefore, that a proper understanding of
experimentally determined phase diagrams of globular proteins should take
into account the nonuniformity of their surfaces and consequently of their interactions.

%%%%%%%%%%%%%%%%%%%%%%%%%%%%%%%%%%%%%%%%%%%%%%%%%%%%%%%%%%%%%%%%%%%%%%%%%%%%%%%
\section[]{Percolation}
\label{sect:percolation}
%%%%%%%%%%%%%%%%%%%%%%%%%%%%%%%%%%%%%%%%%%%%%%%%%%%%%%%%%%%%%%%%%%%%%%%%%%%%%%%
A further interesting issue, already discussed in the context of the
isotropic model, is the percolation threshold,\cite{Fantoni05a,Miller03} to
which we now turn.
%%%%%%%%%%%%%%%%%%%%%%%%%%%%%%%%%%%%%%%%%
\subsection{Virial expansion}
%%%%%%%%%%%%%%%%%%%%%%%%%%%%%%%%%%%%%%%%
In one-patch systems only dimers can form for $\delta$ up to $\pi/6$, while
clusters of any size are in principle possible above this threshold.
In order to study the percolation threshold we can use the strategy
devised in Ref.~\onlinecite{Alon90}.  Based on the definition of the connectedness correlation
function (see below), the percolation threshold is signaled by the divergence of
the mean cluster size
\bq
S=1+\rho\int d\rrr_{12} \left \langle h^+(\rrr_{12},\Omega_1,\Omega_2)
\right \rangle_{\Omega_1,\Omega_2}~,
\label{cluster_size}
\eq
where $h^+$ is the pair connectedness function, which is related to the
direct connectedness function $c^+$ by the Ornstein-Zernike equation.
Both are related to the connected part of the Mayer function $f^{+}(1,2)=f(1,2)-f_{\rm HS}(1,2)$ 
as given in Eq.~(\ref{mayer}), $f_{\rm HS}(1,2)$ being the HS part as given
by Eq.~(\ref{mayer_HS}).
As in the case of Eq.~(\ref{OMF}) we assume that  
\bq
\langle\hat{c}^+(\kkk,\Omega_1,\Omega_3)
\hat{h}^+(\kkk,\Omega_3,\Omega_2)\rangle_{\Omega_3}\simeq
\langle\hat{c}^+(\kkk,\Omega_1,\Omega_3)\rangle_{\Omega_3}
\langle\hat{h}^+(\kkk,\Omega_3,\Omega_2)\rangle_{\Omega_3}~.
\eq
The average Fourier transform of the direct connectedness function
$\hat{c}^+(k)= \langle \hat{c}^+(k,\Omega_1,\Omega_2) 
\rangle_{\Omega_1,\Omega_2}$ at $k=0$ then identifies the threshold by the equation
\bq
\rho \hat{c}^+(0)=1~.
\eq
Upon power expansion in the density we have 
\bq
\hat{c}^+(0)=\sum_{n=2}^\infty\hat{c}^+_n(0)\rho^{n-2} = \hat{c}_2^{+}\left(0\right)+
\rho \hat{c}_3^{+}\left(0\right) + O\left(\rho^2\right).
\eq
Using the above decomposition of the Mayer function the first two coefficients are 
found to be
\bq
\hat{c}^+_2\left(0\right)=\int d\rrr_{12} \left \langle f^+(1,2) \right 
\rangle_{\Omega_1,\Omega_2},
\eq
\begin{multline}
\hat{c}^+_3\left(0\right)=\int d\rrr_{12}d\rrr_{13}
\big \langle \big[
f^+(1,2)f_{\rm HS}(1,3)f_{\rm HS}(2,3)+3f_{\rm HS}(1,2)f^+(1,3)f^+(2,3)+\\
f^+(1,2)f^+(1,3)f^+(2,3) \big] \big \rangle_{\Omega_1,\Omega_2,\Omega_3}.
\end{multline}
An analysis following that in Appendix \ref{app:ve} then yields:
\bq
\hat{c}^+_2(0)/v_0&=&24\frac{\chi_1}{12\tau}~,\\
\hat{c}^+_3(0)/v_0^2&=&60\frac{\chi_1}{12\tau}
-432\frac{\chi_2}{(12\tau)^2}+288\frac{\chi_3}{(12\tau)^3}~.
\eq
To first order in the density, the percolation threshold is then given by a straight line
\bq
\tau=2\chi_1\eta~.
\eq
The next order already yields a more complex solution involving both $\chi_2$ and $\chi_3$
\bq \label{perde}
\eta=\frac{-6\chi_1\tau^2+\sqrt{6}\tau^{3/2}
\sqrt{\chi_3+6\chi_1^2\tau-18\chi_2\tau+30\chi_1\tau^2}}
{\chi_3-18\chi_2\tau+30\chi_1\tau^2}.
\eq
We then see that for $\eta_-\le\eta\le\eta_+$ with
\bq
\eta_\pm=\frac{-6(\chi_1^2-3\chi_2)\pm
\sqrt{36(\chi_1^2-3\chi_2)^2-120\chi_1\chi_3}}{60\chi_1^2}~,
\eq
Eq.~(\ref{perde}) has no real solutions.  Clearly, the acceptable part of the
solution is that for $\eta\le\eta_-$.

%%%%%%%%%%%%%%%%%%%%%%%%%%%%%%%%%%%%%%%%%%%%%%%%%%%%%%%%%%%%%%%%%%%%%%%%%%%%%%%
\subsection[]{C1 integral equation}
%%%%%%%%%%%%%%%%%%%%%%%%%%%%%%%%%%%%%%%%%%%%%%%%%%%%%%%%%%%%%%%%%%%%%%%%%%%%%%%
The unphysical loss of the solution for the percolation threshold as obtained from the virial expansion 
is present also in the isotropic SHS model ($\delta=\pi$).
This shortcoming does not occur in an integral
equation approach.\cite{Fantoni05a}  Within the Cn class of closures
a crucial role is played by the angular average
of the cavity function at contact $\bar{\bar{y}}=\langle y(r_{12}=\sigma,\Omega_1,\Omega_2,
\Omega_{12})\epsilon^2(1,2) \rangle_{\Omega_1,\Omega_2,\Omega_{12}}$. Its
density expansion reads
\bq
\bar{\bar{y}}^{Cn}=y_0+y_1\eta+y_2\eta^2+\ldots~,
\eq
where $y_n=y_n(\tau)$ are related to the reduced virial coefficients $b_n$
by Eq.~(\ref{yn_bn}). For $y_0$ and $y_1$ they have already been computed in Eq. (\ref{y01}).
The percolation threshold is then given by $\eta \bar{y}^{Cn}=\tau$
where $\bar{y}^{Cn}=\langle y^{Cn}(r_{12}=\sigma,\Omega_1,\Omega_2,\Omega_{12})
\epsilon(1,2) \rangle_{\Omega_1,\Omega_2,\Omega_{12}}$
and $y^{Cn}(r_{12}=\sigma,\Omega_1,\Omega_2,\Omega_{12})$ is the
contact cavity function within the Cn approximation. 

\begin{figure}[ht!]
\begin{center}
\includegraphics[width=10cm]{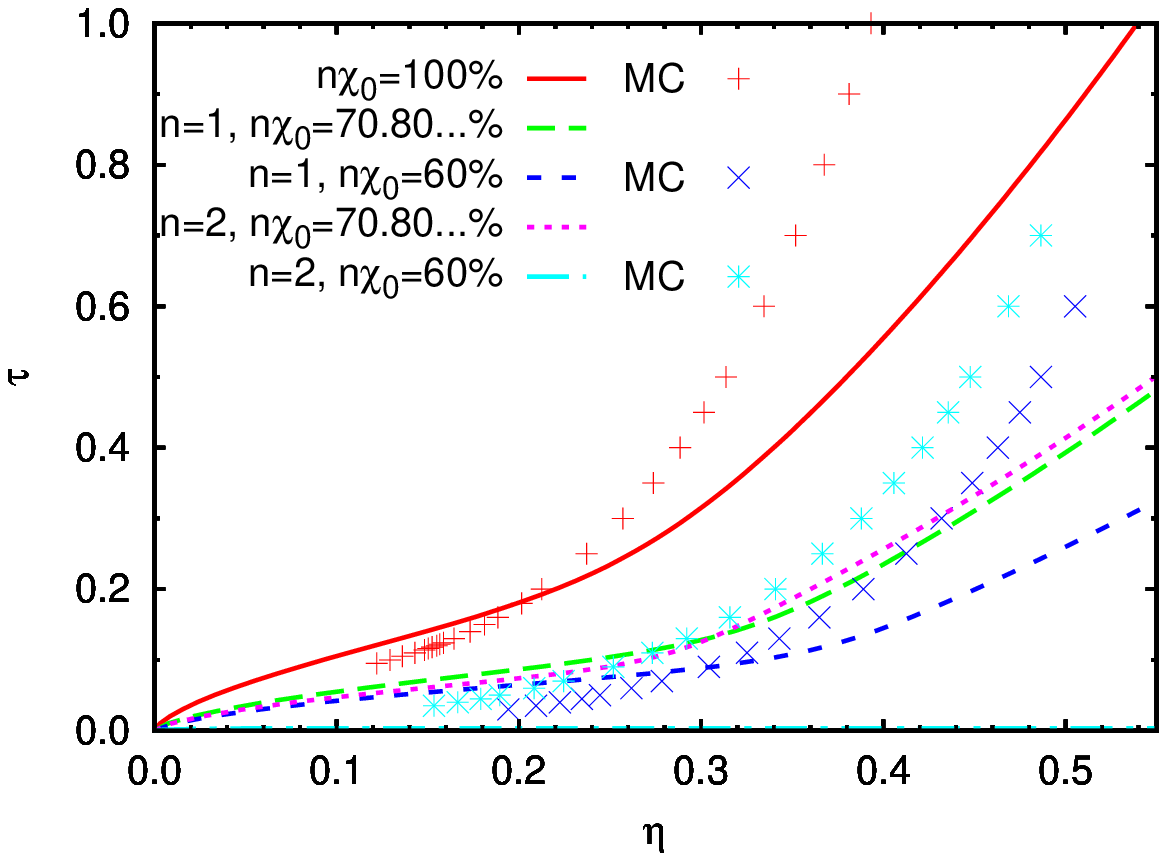}
\end{center}
\caption{}  
\label{fig:percolationC1}
\end{figure}

Since $\epsilon^2=\epsilon$ then $\bar{\bar{y}}=\bar{y}$ and within
the C0 approximation ($\bar{y}^{C0}=y_0$) we find, 
\bq
\tau=\chi_1\eta~,
\eq
whereas within the C1 approximation ($\bar{y}^{C1}=y_0+y_1\eta$) we
find, 
\bq \label{perc1}
\eta=\frac{-6\chi_1\tau^2+\sqrt{12}\tau^{3/2}
\sqrt{\chi_3+3\chi_1^2\tau-12\chi_2\tau+30\chi_1\tau^2}}
{\chi_3-12\chi_2\tau+30\chi_1\tau^2}~.
\eq
Now the loss of solution occurs between
\bq
\eta_\pm=\frac{-3(\chi_1^2-4\chi_2)\pm
\sqrt{9(\chi_1^2-4\chi_2)^2-120\chi_1\chi_3}}{30\chi_1^2}~.
\eq
Note that at small values of $\delta$ a gap may also appear in
the C1 percolation threshold for
$\delta\lesssim 1.21$ in the one-patch model and for
$\delta\lesssim 1.22$ in the two-patch case. 
Fig. \ref{fig:percolationC1} summarizes our findings and compares
with MC simulations.  From
the figure we see that for $\delta$ close to $\pi$ the
percolation threshold of the two-patch model lies
above that of the one-patch case at same total surface
adhesive coverage, while the opposite trend is observed at lower
$\delta$.  This mirrors our previous results for the coexistence
curve.

\begin{figure}[ht!]
\begin{center}
\includegraphics[width=10cm]{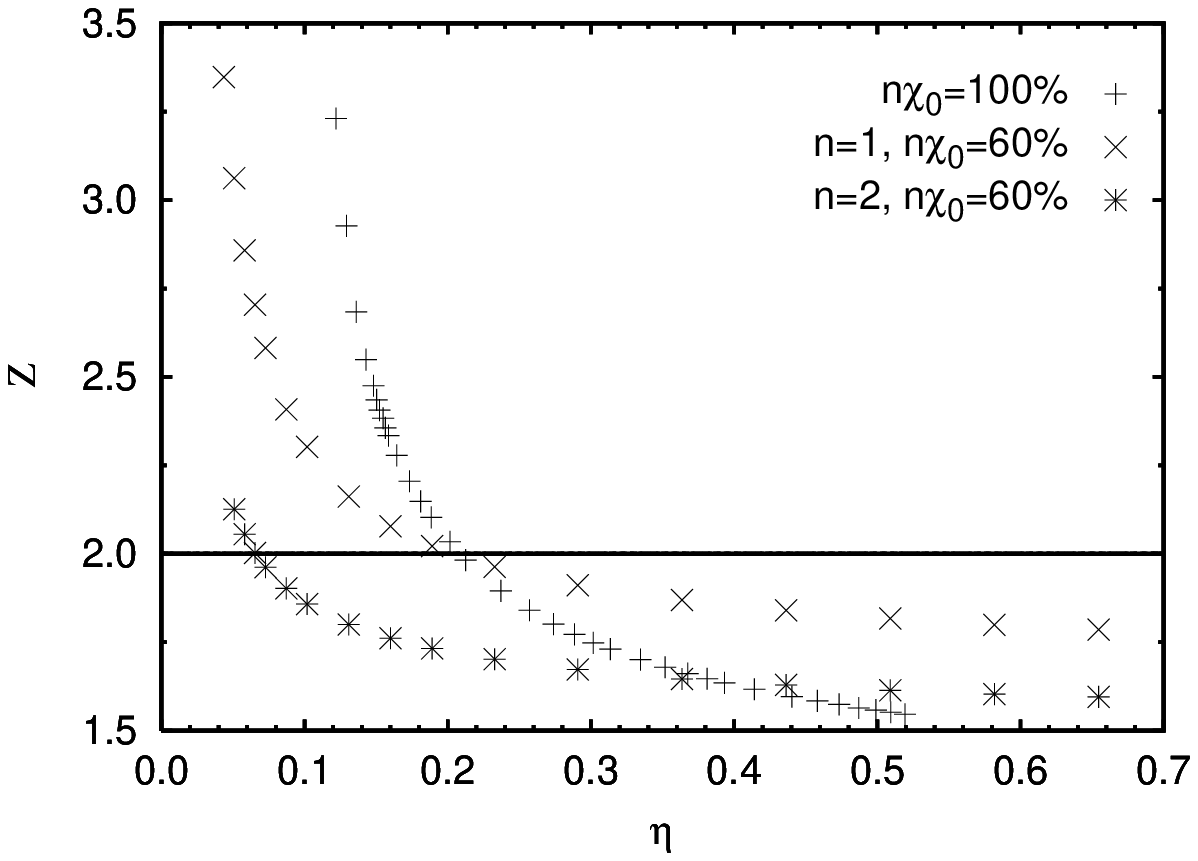}
\end{center}
\caption{}
\label{fig:coord}
\end{figure}

Another quantity which is useful to assess the onset of a phase transition
is the average coordination number, defined by
\bq
Z &=& \rho \int d\rrr_{12} \left\langle h^{+}\left(1,2\right) 
\right \rangle_{\Omega_1,\Omega_2,\Omega_{12}}.
\label{averageZ}
\eq
One finds
\bq
Z=2\frac{\eta}{\tau}\bar{y}^{Cn}~,
\eq
which on the percolation threshold gives $Z=2$. This prediction is
compared with MC results in
Fig.~\ref{fig:coord} where we show the average coordination number at the
percolation threshold obtained from the MC simulations for the
one- and two-patch models at $60\%$ coverage.

\begin{figure}[ht!]
\begin{center}
\includegraphics[width=10cm]{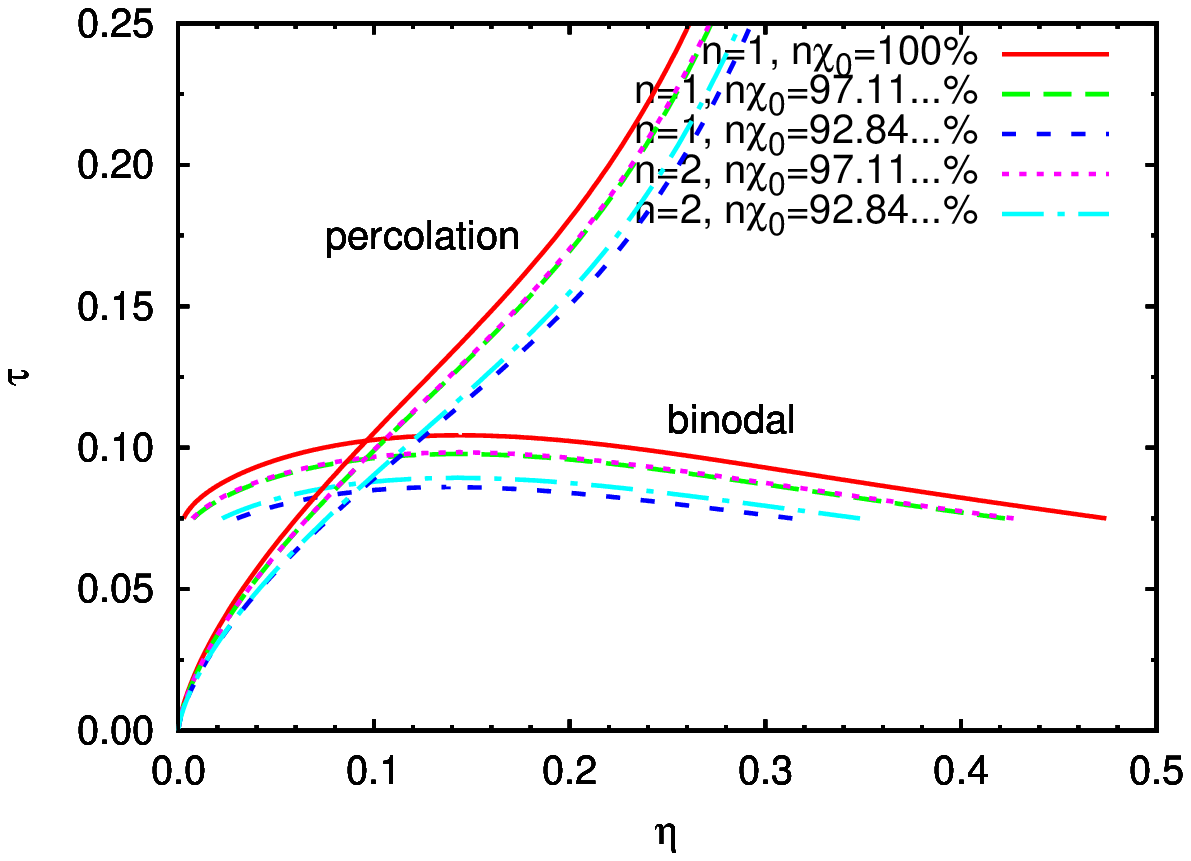}
\end{center}
\caption{}
\label{fig:pd-C1}
\end{figure}

\begin{figure}[ht!]
\begin{center}
\includegraphics[width=10cm]{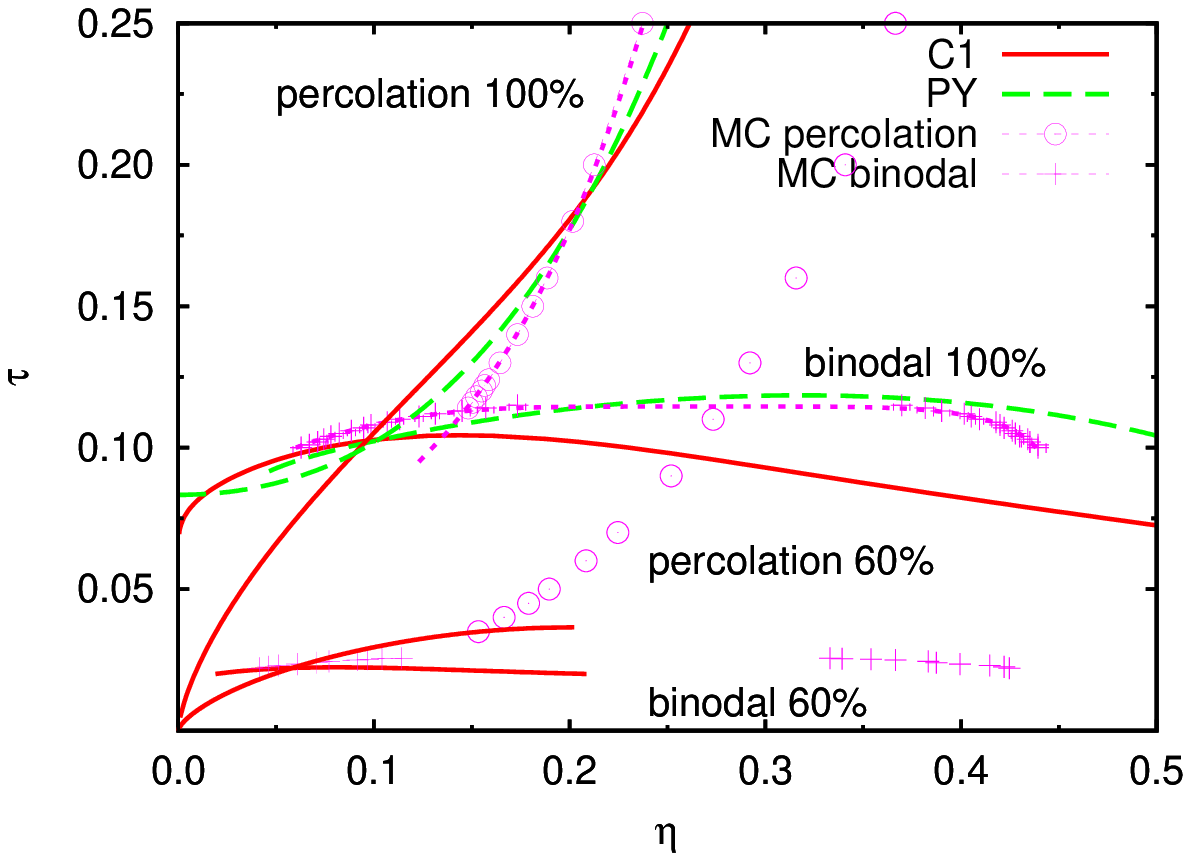}
\end{center}
\caption{}
\label{fig:pd}
\end{figure}

We are now in a position to summarize the phase diagram for one
and two patches within the C1 approximation. This is reported in
Fig.~\ref{fig:pd-C1}.
For two patches, the C1 phase diagram (coexistence curve and percolation line) is
compared with MC results in Fig.~\ref{fig:pd} both for the full isotropic case
($100\%$ coverage) and for $60\%$ coverage.  Note that while the percolation line
terminates at the point shown, the coexistence curve has a solution for the
whole range of packing fraction considered.  However, we have chosen to
terminate the plot for the same value of $\eta$ as the percolation line.

%%%%%%%%%%%%%%%%%%%%%%%%%%%%%%%%%%%%%%%%%%%%%%%%%%%%%%%%%%%%%%%%%%%%%%%%%%%%%%%
\section{Phase Diagram and addition of adhesive background}
\label{sect:background}
%%%%%%%%%%%%%%%%%%%%%%%%%%%%%%%%%%%%%%%%%%%%%%%%%%%%%%%%%%%%%%%%%%%%%%%%%%%%%%%

So far we have considered the case of adhesive patches on hard spheres.
The disadvantage
of this model is that there is no fluid-fluid transition below a certain surface coverage $\chi_0$. 
We have also shown that at fixed surface coverage the liquid more easily forms if the 
adhesion is distributed in different patches on the sphere surface for sufficiently
large patches and we expect the opposite to be true for low coverage.

One could argue that a more physical model should have a strongly directional
potential mimicking e.g., active sites in a globular protein, in addition to an
underlying isotropic attractive potential favoring a general fluid-fluid
phase transition.
To this aim, we modify our potential by adding a uniform adhesive background to
each sphere on top of which a patchy potential of the type considered so far is active.
This effect can be obtained by a simple substitution
$\epsilon\rightarrow 1+\lambda\epsilon$ with $\lambda$ measuring the strength
of adhesion on the patches, yielding
\bq
\chi_1&\rightarrow&1+\lambda\chi_1~,\\
\chi_2&\rightarrow&1+2\lambda\chi_1+\lambda^2\chi_2~,\\
\chi_3&\rightarrow&1+3\lambda\chi_1+3\lambda^2\bar{\chi}_2+\lambda^3\chi_3~,
\eq
where 
\bq
\bar{\chi}_2=\langle\epsilon(1,2)\epsilon(1,3)\rangle_{\Omega_1,\Omega_2,\Omega_3}|_{\theta_{12}=\pi/3}~.
\eq

\begin{figure}[ht!]
\begin{center}
\includegraphics[width=8cm]{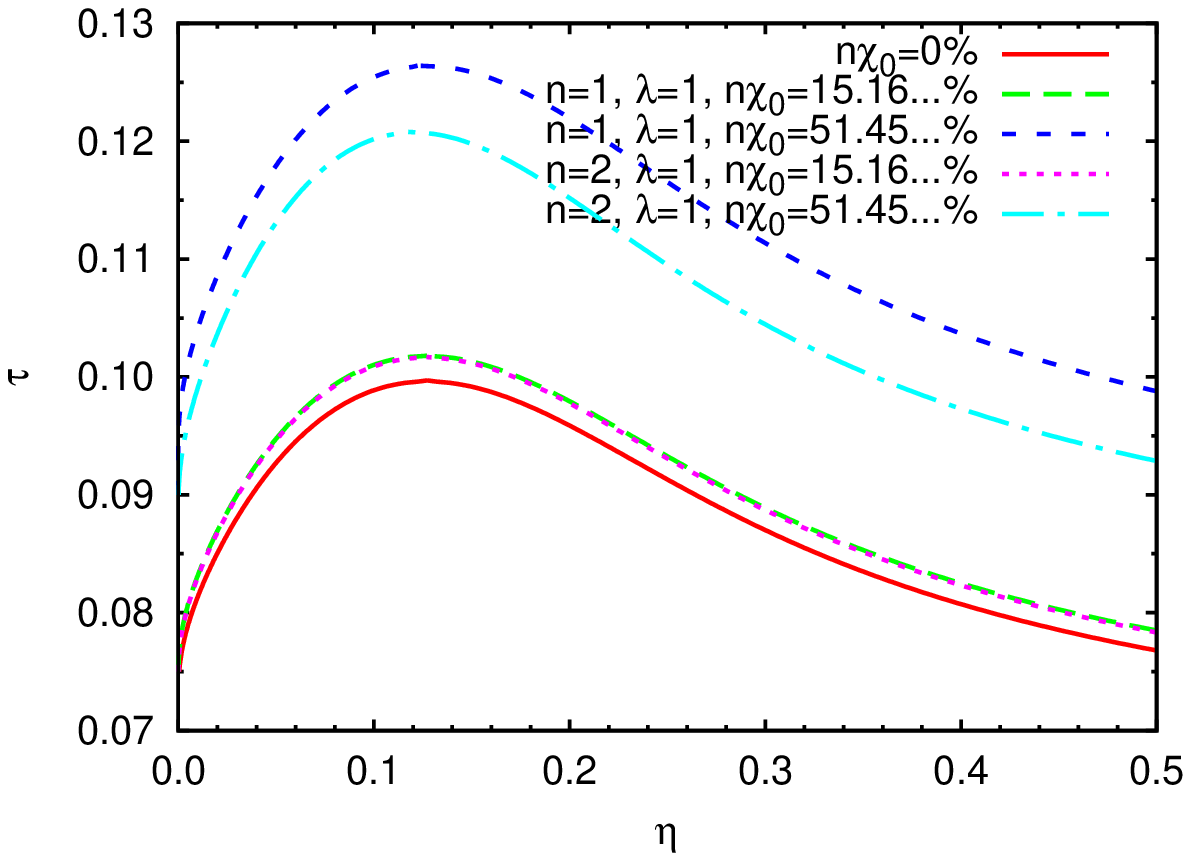}
\includegraphics[width=8cm]{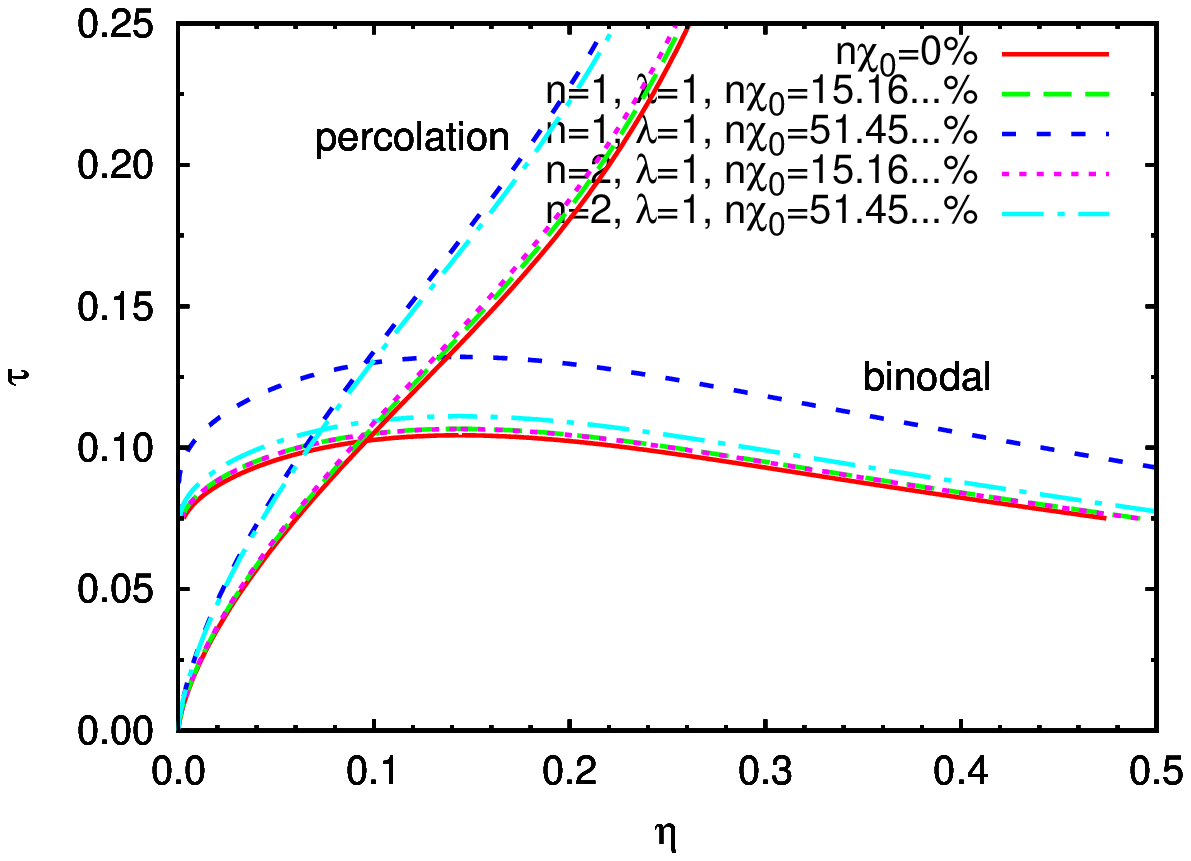}
\end{center}
\caption{}
\label{fig:binodalb}
\end{figure}

The phase diagram is now modified as depicted in Fig.~\ref{fig:binodalb} where
we have set $\lambda=1$ to be the strength of the patches throughout.
In this case we see that even a small sticky patch (of amplitude $\delta \sim 0.5$)
is sufficient to raise both the binodal and percolation threshold of the isotropic model. 
At equal coverage, the binodal and percolation threshold of the two-patch model
lie below their one-patch counterparts, in agreement with the observed trend
in the absence of background adhesion.

\begin{figure}[ht!]
\begin{center}
\includegraphics[width=10cm]{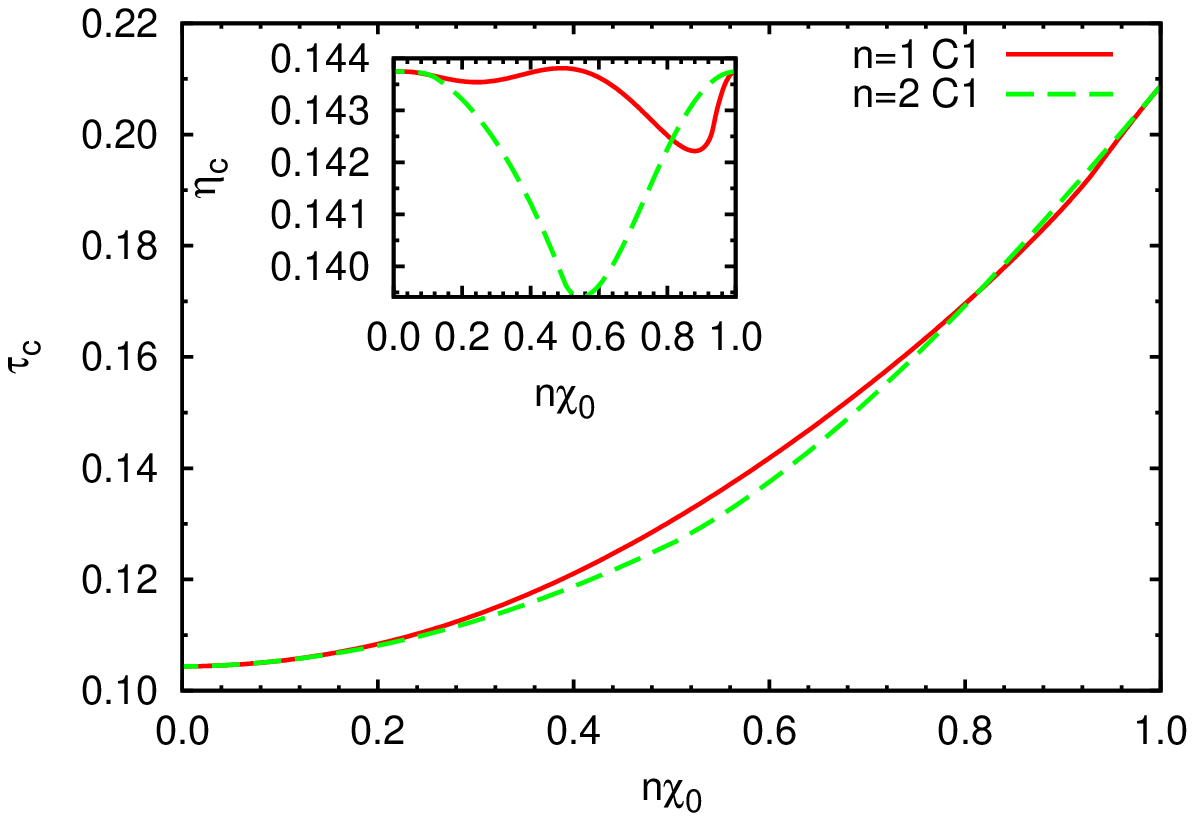}
\end{center}
\caption{}
\label{fig:criticalb1c1}
\end{figure}

Note that the critical point is now less sensitive to the size of the
patches because an isotropic SHS---rather than a hard sphere---is
now the limiting case as $\delta\to0$. Indeed, the critical point does 
not move along $\eta$ while it 
covers the whole range $\tau_c<\tau<2\tau_c$ ($\lambda=1$), where $\tau_c$
is the critical reduced temperature of the isotropic model (see
Fig.~\ref{fig:criticalb1c1}). The critical point shifts of the one- and
two-patch models are now almost indistinguishable even
though the crossing at $80\%$ coverage still remains.

%%%%%%%%%%%%%%%%%%%%%%%%%%%%%%%%%%%%%%%%%%%%%%%%%%%%%%%%%%%%%%%%%%%%%%%%%%%%%%%
\section{Conclusions}
\label{sect:conclusions}
%%%%%%%%%%%%%%%%%%%%%%%%%%%%%%%%%%%%%%%%%%%%%%%%%%%%%%%%%%%%%%%%%%%%%%%%%%%%%%%

In this work we have studied, through integral equation theories and Monte
Carlo simulations, the structure, percolation and fluid-fluid
coexistence curves of a model of hard spheres with one or two
uniform sticky patches on their surface.  Particles interact through an adhesive Baxter
potential only if patches on different spheres are suitably oriented and
as hard spheres otherwise. 
Unlike most previous studies, we have been able to analyze in some detail
the dependence of the aforementioned quantities on the size
of the patch and its interplay with the number of patches.

The integral equation theory
is based on the first two approximations (C0 and C1) of a class of closures Cn
which have already proved to provide a good qualitative representation
of the exact behavior and are almost fully analytical. 
The comparison between the
analytical work and Monte Carlo simulations indicates that C1 yields
a gratifying qualitative description of the phase diagram
notwithstanding the expected limitation due to its low density nature.
While for the thermodynamics the results from the integral equation
theories are exact within the given closure, for the percolation
problem and the structure an 
additional orientational mean field approximation is necessary to decouple
the orientational average.

Radial distribution functions within the C0 and C1 integral
equation theories exhibit a characteristic jump at $r=2\sigma$ (whose magnitude
depends on the patch angle $\delta$) and a cusp at $r=3\sigma$.
The coexistence and percolation lines move to lower temperature as the patch angle
decreases from $\pi$ (the isotropic case) to $0$ (hard spheres). 
For a fixed surface coverage above approximately $80\%$, the curves of the
two-patch case lie above the corresponding single-patch ones, while
the opposite trend is observed below that point.
We have suggested that this is due to two patches
of sufficiently large size being able to form bonds to more particles than can a
single patch. We have also argued that this reasoning does not apply at low
coverage, and that, in fact, the opposite situation might be expected. 
The crossover is not observed in the MC simulations within the range of adhesive
coverage studied here (about $60\%$), but we cannot exclude the possibility
for lower coverage, where the simulations converge very slowly.
When an adhesive background is included in addition to the patches, both the liquid and
percolating phase of the system are favored with respect to the isotropic case  
even in the presence of very small patches.

In spite of the limited number of cases (one or two patches) addressed in
the present work, our analysis suggests that both
the total fraction of the surface covered by adhesion and the number
of patches are crucial
parameters in controlling the location of the critical point. In the limit 
of a single bond per patch, our analysis is consistent with a recent 
suggestion\cite{Foffi07} of a generalized law of corresponding states for 
non-isotropic patchy interactions. 
We remark that, from the purely theoretical point of view, there
exist only few paradigmatic toy models
with anisotropic interactions amenable to analytical or semi-analytical treatment.

Our analysis can be regarded as complementary to recent investigations
of the phase diagrams of globular proteins\cite{tenWolde97,Sear99,Lutsko05}
in that our starting point is the isotropic sticky hard sphere from which
some adhesion is removed, rather that a hard sphere to which highly localized attractive
spots are added.  This approach goes beyond the limitation of
one bond per patch, which is an essential feature of Wertheim thermodynamic
perturbation theory.  The price to pay is, of course, that only
a qualitative agreement with MC simulations can be achieved. 

It would be interesting to extend the present work in some respects.
In view of the difficulties of MC simulations
in probing low coverage, a comparison with a numerical solution
of a more robust closure such as, for instance, 
the Percus-Yevick approximation which has a full analytical description
in the isotropic case, would provide a more quantitative
assessment of the results presented here.  Such a solution would also
help to evaluate the
(uncontrolled) angular decoupling approximation exploited
in the present analysis of structure and the percolation threshold.
Work along these lines is in progress and will be presented in a future publication.

%%%%%%%%%%%%%%%%%%%%%%%%%%%%%%%%%%%%%%%%%%%%%%%%%%%%%%%%%%%%%%%%%%%%%%%%%%%%%%%
\begin{acknowledgments}
This work was supported by the Italian MIUR (PRIN-COFIN 2006/2007).
MAM thanks the Royal Society of London and EPSRC for financial support.
\end{acknowledgments}
%%%%%%%%%%%%%%%%%%%%%%%%%%%%%%%%%%%%%%%%%%%%%%%%%%%%%%%%%%%%%%%%%%%%%%%%%%%%%%%

\appendix
%%%%%%%%%%%%%%%%%%%%%%%%%%%%%%%%%%%%%%%%%%%%%%%%%%%%%%%%%%%%%%%%%%%%%%%%%%%%
\section{The law of corresponding states}
\label{app:lcs}
%%%%%%%%%%%%%%%%%%%%%%%%%%%%%%%%%%%%%%%%%%%%%%%%%%%%%%%%%%%%%%%%%%%%%%%%%%%
Consider the simplest possible dependence $\Delta(\hat{\sss}_1, \hat{\sss}_2)= \hat{\sss}_1 \cdot
 \hat{\sss}_2 $ and assume that 
$\epsilon(1,2)=\Delta(\hat{\sss}_1,\hat{\sss}_2)$, i.e. the adhesion coefficient
does not depend on $\hat{\rrr}_{12}=\rrr_{12}/r_{12}$. Within the 
Weeks-Chandler-Andersen perturbative expansion\cite{Andersen71} 
of the Helmholtz free energy $A^{\rm SHS}$ one finds
\begin{multline}
\frac{\beta(A^{\rm SHS}-A^{\rm HS})}{N}=\int d(1) d(2) \,a^{(1)}(\rrr_1,\rrr_2;\eta)
\Delta e(1,2)+\\
\int d(1) d(2) d(3) d(4) \,a^{(2)}(\rrr_1,\rrr_2,\rrr_3,\rrr_4;\eta)
\Delta e(1,2)\Delta e(3,4)+\ldots
\end{multline}
where $d(i)$ is a short-hand notation for $d\rrr_id\tilde{\Omega}_i$, 
with $d\tilde{\Omega}_i$ the average solid angle 
$\sin\theta_id\theta_i\,d\varphi_i/(4\pi)$, $A^{\rm HS}$ the Helmholtz
free energy of the reference hard sphere (HS) system, 
\bq
\Delta e(1,2)=\frac{\epsilon(1,2)}{12\tau}\delta(r_{12}-\sigma)~,
\eq
and the functions $a^{(n)}$ are expressed in terms of the correlation 
functions of the reference system which only depend on the packing 
fraction $\eta=\pi\rho\sigma^3/6$, with $\rho$ the density. We see then
that the angular dependence in $\epsilon$ factorizes and one finds
\bq
\frac{\beta(A^{\rm SHS}-A^{\rm HS})}{N}=\sum_iA^{(i)}(\eta)
\left(\frac{\chi_1}{12\tau}\right)^i~,
\eq
where $\chi_1=\int
d\tilde{\Omega}_1d\tilde{\Omega}_2\,\epsilon(1,2)$.  This analysis shows how,
in this case, the {\em law of corresponding states} holds.  For 
example, if $\tau=g(\eta)$ is the spinodal or binodal of the SHS system
with isotropic interaction $(\epsilon=1)$,\cite{Baxter68,Baxter71,Watts71}
then the spinodal or binodal of the SHS with directional adhesion will be
$\tau=\chi_1 g(\eta)$, which will lie above that of the isotropic 
system if $\chi_1>1$ and below otherwise.
%%%%%%%%%%%%%%%%%%%%%%%%%%%%%%%%%%%%%%%%%%%%%%%%%%%%%%%%%%%%%%%%%%%%%%
\section{The 3rd virial coefficient}
\label{app:ve}
%%%%%%%%%%%%%%%%%%%%%%%%%%%%%%%%%%%%%%%%%%%%%%%%%%%%%%%%%%%%%%%%%%%%%%
In this appendix we provide a derivation of Eqs.~(\ref{virial}), (\ref{chi1}),
(\ref{chi2}), and (\ref{chi3}). We start from the usual definition of the
third virial coefficient
\bq
B_3 &=& -\frac{1}{3V} \int d\rrr_1~d\rrr_2 ~d\rrr_3 \left \langle f\left(1,2\right)
f\left(1,3\right) f\left(2,3\right) \right \rangle_{\Omega_1,\Omega_2,\Omega_3}
\label{3rd}
\eq
where, in line with Eq.~(\ref{SHS}), the Mayer function can split into two terms
\bq
f\left(i,j\right) &=& f_{\rm HS}\left(i,j\right) + \frac{\sigma}{12 \tau} \epsilon\left(i,j\right)
\delta\left(r_{ij}-\sigma\right).
\label{mayer}
\eq
In the above equation we have set the HS part to the usual form
\bq
f_{\rm HS}\left(i,j\right) &=& - \Theta\left(\sigma - r_{ij} \right).
\label{mayer_HS}
\eq
Upon expanding the product, one can easily find 
\bq
\Delta B_{3}= B_3-B_3^{\rm HS} = \Delta B_3^{(1)}+\Delta B_3^{(2)}+\Delta B_3^{(3)},
\label{deltaB}
\eq
where $B_3^{\rm HS}=5 \pi^2 \sigma^6/18$ is the HS result and
\bq
\Delta B_{3}^{(1)} &=& - \frac{1}{V} \left(\frac{\sigma}{12 \tau}
\right)
\int d\rrr_1~d\rrr_2 ~d\rrr_3 \left \langle f_{\rm HS}\left(1,2\right)
f_{\rm HS}\left(1,3\right) \epsilon\left(2,3\right) \delta\left(r_{23}-\sigma\right) 
\right \rangle_{\Omega_1,\Omega_2,\Omega_3}
\label{deltaB1}
\eq
\begin{multline}
\Delta B_{3}^{(2)} = - \frac{1}{V} \left(\frac{\sigma}{12 \tau}
\right)^2
\int d\rrr_1~d\rrr_2 ~d\rrr_3\\
\left \langle f_{\rm HS}\left(1,2\right)
\epsilon\left(1,3\right) \delta\left(r_{13}-\sigma\right)
\epsilon\left(2,3\right) \delta\left(r_{23}-\sigma\right) 
\right \rangle_{\Omega_1,\Omega_2,\Omega_3}
\label{deltaB2}
\end{multline}
\begin{multline}
\Delta B_{3}^{(3)} = - \frac{1}{3V} \left(\frac{\sigma}{12 \tau}
\right)^3 
\int d\rrr_1~d\rrr_2 ~d\rrr_3\\
\left \langle 
\epsilon\left(1,2\right) \delta\left(r_{12}-\sigma\right)
\epsilon\left(1,3\right) \delta\left(r_{13}-\sigma\right)
\epsilon\left(2,3\right) \delta\left(r_{23}-\sigma\right) 
\right \rangle_{\Omega_1,\Omega_2,\Omega_3}.
\label{deltaB3}
\end{multline}
The above integrals are most conveniently evaluated in bipolar
coordinates by introducing $\rrr_{12} =\rrr_{2}-\rrr_{1}$ and
$\rrr_{13}=\rrr_{3}-\rrr_{1}$. 
This leads to $r_{23}=\sqrt{r_{12}^2+r_{13}^2-2
  r_{12}r_{13}\hat{\rrr}_1 \cdot \hat{\rrr}_2}$  
where it is most convenient to choose $\hat{\rrr}_{13}$ as the $z$
axis. For $\Delta B_3^{(1)}$ one finds
\bq
\Delta B_{3}^{(1)} &=& - \left(\frac{\sigma}{12 \tau} \right)
\int d\rrr_{23} d\rrr_{12} \Theta\left(\sigma-r_{12}\right)
\Theta\left(\sigma-|\rrr_{12}-\rrr_{23}|\right)\delta\left(r_{23}-\sigma\right)
\left \langle \epsilon\left(2,3\right)
\right \rangle_{\Omega_2,\Omega_3}
\label{deltaB11}
\eq
Here one first performs the integration over $\rrr_{12}$, which covers
twice a spherical cap of height $\sigma/2$ and then the
straightforward integration over $\rrr_{23}$.
Clearly the anisotropic part decouples, thus yielding the isotropic
part times $\chi_1$ as claimed. For $\Delta B_3^{(2)}$ a little more
care is necessary. One first obtains
\begin{multline}
\Delta B_3^{(2)} = \left(\frac{\sigma}{12 \tau}\right)^2 \int_{0}^{\infty} d
r_{12} r_{12}^2 
\Theta\left(\sigma - r_{12}\right) \int_{0}^{\infty} d r_{13} r_{13}^2 
\delta\left(\sigma -r_{13} \right) \\
\int d \Omega_{12} d\Omega_{13}
\delta\left(\sqrt{r_{12}^2+r_{13}^2-2 r_{12} r_{13} \cos \theta_{12}} -\sigma\right)
\left \langle \epsilon\left(1,3\right) \epsilon\left(2,3\right) 
\right \rangle_{\Omega_1,\Omega_2,\Omega_3}.
\label{deltaB21}
\end{multline}
After a first integration over $r_{13}$, an additional integration over $\cos \theta_{12}$ 
then requires $\theta_{13} \le \pi/3$
(corresponding to the maximum available angle for all three particles in reciprocal contact).
This also yields a normalization factor $4=1/\sin^2(\pi/6)$ in order to have the correct
limit $\epsilon(i,j) \to 1$ for all $(i,j)$. The final result is
\bq
\Delta B_3^{(2)} &=& \frac{\pi^2 \sigma^6}{36 \tau^2} \frac{1}{4\pi} \int d\Omega_{13}
~4~\left \langle \epsilon\left(1,2\right) \epsilon\left(2,3\right) 
\theta\left(\frac{\pi}{3}-\theta_{13}\right) \right \rangle_{\Omega_1,\Omega_2,\Omega_3},
\label{deltaB22}
\eq
thus yielding the isotropic part times $\chi_2$ as reported in Eq.~(\ref{chi2}). An almost
identical procedure also gives 
\begin{multline}
\Delta B_3^{(3)} = - \left(\frac{\sigma}{12\tau}\right)^3 
\frac{8\pi^2\sigma^2}{3}\int_{0}^{\infty} 
d r_{12} r_{12}^2 \delta\left(r_{12}-\sigma\right)\int_{-1}^{+1} d
\left(\cos \theta_{12}\right) \\ 
\delta \left(\sqrt{r_{12}^2+\sigma^2-2 r_{12}\sigma \cos
\theta_{12}} -\sigma \right) 
\left \langle \epsilon\left(1,2\right) \epsilon\left(1,3\right)
\epsilon\left(2,3\right) 
\right \rangle_{\Omega_1,\Omega_2,\Omega_3}
\label{deltaB31}
\end{multline}
which, after an integration over the angular variables, leads to the desired decoupling
for the anisotropic part $\chi_3$ as given in Eq.~(\ref{chi3}). Note that in this
configuration all three spheres are necessarily touching and this fixes the angles
$\theta_{ij}$ to a well defined value given in Eq.~(\ref{chi3}). 
This completes the derivation of Eq.~(\ref{virial}).
%%%%%%%%%%%%%%%%%%%%%%%%%%%%%%%%%%%%%%%%%%%%%%%%%%%%%%%%%%%%%%%%%%%%%%%%%%%%%%%%%%%%%%%%%%%
\section{Coefficients $\chi_2$ and $\chi_3$  for the one-patch case}
\label{app:op}
\begin{figure}[ht!]
\begin{center}
\includegraphics[width=10cm]{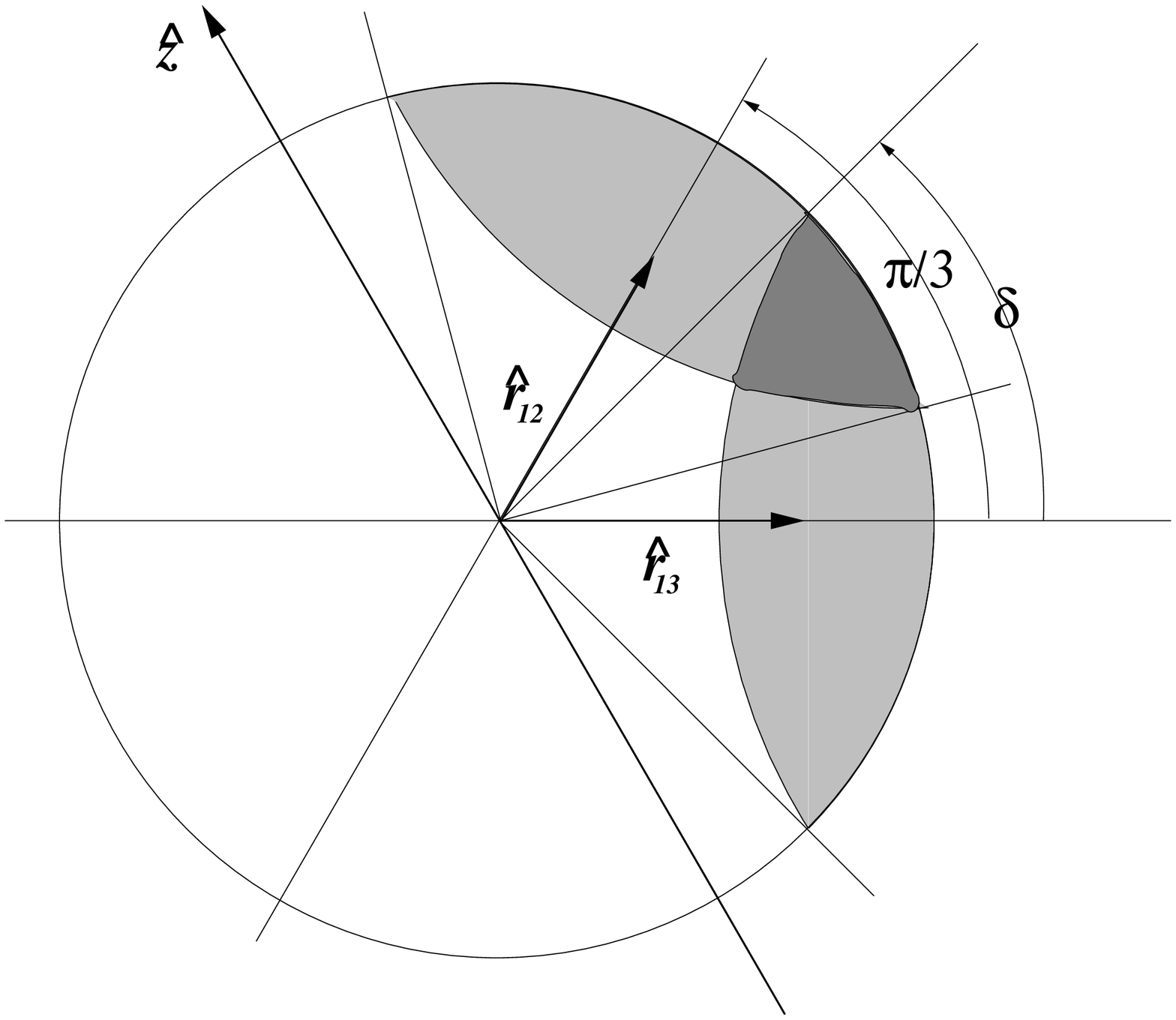}
\end{center}
\caption{}
\label{fig:cones}
\end{figure}

%%%%%%%%%%%%%%%%%%%%%%%%%%%%%%%%%%%%%%%%%%%%%%%%%%%%%%%%%%%%%%%%%%%%%%%%%%%%%%%%%%%%%%%%%%%
Here we give the analytic expressions for the coefficients $Q_1$ and
$R_1$ used in Eqs.~(\ref{1c2}) and (\ref{1c3}) of the main text in terms
of characteristic integrals which are then evaluated numerically.
The basic procedure follows a similar analysis carried out
in a different context,\cite{Oversteegen03} which requires the calculation
of the solid angle associated with the intersection of two identical patches
on the same sphere as indicated in Fig.~\ref{fig:cones}.
For $\chi_3$, one can easily see that for $\delta < \pi/6$ there is no possibility of
intersection, even in the close-packed configuration. For $\delta \ge \pi/6$
the form of the resulting integral can be most conveniently written in
slightly different ways depending on the amplitude $\delta$ of the patch.
\begin{multline}
R_1(\delta)=\alpha_{b,1}(\delta)\Theta\left(\delta-\frac{\pi}{6}\right)
\Theta\left(\frac{2\pi}{3}-\delta\right)
+\beta_{b,1}(\delta)\Theta\left(\delta-\frac{2\pi}{3}\right)
\Theta\left(\frac{5\pi}{6}-\delta\right)+\\
\gamma_{b,1}(\delta)\Theta\left(\delta-\frac{5\pi}{6}\right),
\label{Bop}
\end{multline}
where the various terms are given in terms of the integrals
\bq \label{Bop1}
&&\alpha_{b,1}(\delta)=\frac{1}{\pi}\int_{2\pi/3-\delta}^{\pi/2}
d\theta~\sin\theta\arccos\left(\frac{\cos\delta-\cos\theta\cos 2\pi/3}
{\sin\theta\sin 2\pi/3}\right)\,\\ \label{Bop2}
&&\beta_{b,1}(\delta)=1-\frac{1}{\pi}\int_{\delta-2\pi/3}^{\pi/2}
d\theta~\sin\theta\arccos\left(\frac{\cos(\pi-\delta)-\cos\theta\cos \pi/3}
{\sin\theta\sin \pi/3}\right)\,\\ \label{Bop3}
&&\gamma_{b,1}(\delta)=1-2\sin^2\left(\frac{\pi-\delta}{2}\right)~.
\eq
For example $\alpha_{b,1}$ given in Eq.~(\ref{Bop1}) is the simplest integral
resulting from the calculation of the overlapping region of the two
cones of width $\delta$ as depicted in Fig. \ref{fig:cones}.

For $\chi_2$ an additional complication arises from the additional degree
of freedom given by the fact that only two of the three spheres are
(in general) in contact. One finds
\begin{multline}
Q_1(\delta)=\alpha_{a,1}(\delta)\Theta\left(\frac{\pi}{6}-\delta\right)
+\beta_{a,1}(\delta)\Theta\left(\delta-\frac{\pi}{6}\right)
\Theta\left(\frac{\pi}{2}-\delta\right)+\\
\gamma_{a,1}(\delta)\Theta\left(\delta-\frac{\pi}{2}\right)
\Theta\left(\frac{5\pi}{6}-\delta\right)
+\delta_{a,1}(\delta)\Theta\left(\delta-\frac{5\pi}{6}\right),
\label{Ad}
\end{multline}
with
\bq
\alpha_{a,1}(\delta)=&&\frac{2}{\pi}
\int_{0}^{2\delta}d\theta^\prime
\sin\theta^\prime\int_{\pi/2+\theta^\prime/2-\delta}^{\pi/2}
d\theta~\sin\theta\arccos
\left(\frac{\cos\delta-\cos\theta\cos(\pi/2+\theta^\prime/2)}
{\sin\theta\sin(\pi/2+\theta^\prime/2)}\right),\\
\beta_{a,1}(\delta)=&&\frac{2}{\pi}
\int_{0}^{\pi/3}d\theta^\prime
\sin\theta^\prime\int_{\pi/2+\theta^\prime/2-\delta}^{\pi/2}
d\theta~\sin\theta\arccos
\left(\frac{\cos\delta-\cos\theta\cos(\pi/2+\theta^\prime/2)}
{\sin\theta\sin(\pi/2+\theta^\prime/2)}\right),\\
\gamma_{a,1}(\delta)=&&\frac{2}{\pi}
\left[\frac{\pi}{2}-\int_{0}^{\pi/3}d\theta^\prime\sin\theta^\prime\right.\nonumber\\
&&\left.\int_{\delta-\pi/2-\theta^\prime/2}^{\pi/2}
d\theta~\sin\theta\arccos
\left(\frac{\cos(\pi-\delta)-\cos\theta\cos(\pi/2-\theta^\prime/2)}
{\sin\theta\sin(\pi/2-\theta^\prime/2)}\right)\right],\\
\delta_{a,1}(\delta)=&&\frac{2}{\pi}\Bigg\{\frac{\mbox{}}{\mbox{}}
2\pi\sin^2\delta+\int_{0}^{2(\pi-\delta)}d\theta^\prime\sin\theta^\prime\nonumber\\
&&\int_{\delta-\pi/2-\theta^\prime/2}^{\pi/2}
d\theta \sin\theta\arccos
\left(\frac{\cos(\pi-\delta)-\cos\theta\cos(\pi/2-\theta^\prime/2)}
{\sin\theta\sin(\pi/2-\theta^\prime/2)}\right)+\\
&&\left[1-2\sin^2\left(\frac{\pi-\delta}{2}\right)\right]
\frac{\pi}{2}\left[2\cos(2\delta)-1\right]\Bigg\}.
\eq

%%%%%%%%%%%%%%%%%%%%%%%%%%%%%%%%%%%%%%%%%%%%%%%%%%%%%%%%%%%%%%%%
\section{Coefficients $\chi_2$ and $\chi_3$ for the two-patch case}
\label{app:tp}
%%%%%%%%%%%%%%%%%%%%%%%%%%%%%%%%%%%%%%%%%%%%%%%%%%%%%%%%%%%%%%%%
Here we give the analytic expressions for the coefficients $Q_2$ and
$R_2$ used in Eqs.~(\ref{2c2}) and (\ref{2c3}) of the main text:
\bq
Q_2(\delta)&=&\alpha_{a,2}(\delta)\Theta\left(\frac{\pi}{6}-\delta\right)+
\beta_{a,2}(\delta)\Theta\left(\delta-\frac{\pi}{6}\right)
\Theta\left(\frac{\pi}{3}-\delta\right)+\nonumber\\ \label{Ad2}
&&\gamma_{a,2}(\delta)\Theta\left(\delta-\frac{\pi}{3}\right)
\Theta\left(\frac{\pi}{2}-\delta\right)~,\\
R_2(\delta)&=&\alpha_{b,2}(\delta)\Theta\left(\delta-\frac{\pi}{6}\right)
\Theta\left(\frac{\pi}{3}-\delta\right)
+\beta_{b,2}(\delta)\Theta\left(\delta-\frac{\pi}{3}\right)
\Theta\left(\frac{\pi}{2}-\delta\right)~,
\eq
with
\bq
\alpha_{a,2}(\delta)&=&\frac{4}{\pi}
\int_{0}^{2\delta}d\theta^\prime\sin\theta^\prime
\int_{\pi/2+\theta^\prime/2-\delta}^{\pi/2}
d\theta~\sin\theta\arccos
\left(\frac{\cos\delta-\cos\theta\cos(\pi/2+\theta^\prime/2)}
{\sin\theta\sin(\pi/2+\theta^\prime/2)}\right),\\
\beta_{a,2}(\delta)&=&\frac{4}{\pi}
\int_{0}^{\pi/3}d\theta^\prime\sin\theta^\prime
\int_{\pi/2+\theta^\prime/2-\delta}^{\pi/2}
d\theta ~\sin\theta\arccos
\left(\frac{\cos\delta-\cos\theta\cos(\pi/2+\theta^\prime/2)}
{\sin\theta\sin(\pi/2+\theta^\prime/2)}\right),\\ \nonumber
\gamma_{a,2}(\delta)&=&\frac{4}{\pi}\left[
\int_{0}^{\pi/3}d\theta^\prime\sin\theta^\prime
\int_{\pi/2+\theta^\prime/2-\delta}^{\pi/2}
d\theta~\sin\theta\arccos
\left(\frac{\cos\delta-\cos\theta\cos(\pi/2+\theta^\prime/2)}
{\sin\theta\sin(\pi/2+\theta^\prime/2)}\right)+\right.\\
&&\left.\int_{0}^{\pi/3}d\theta^\prime\sin\theta^\prime
\int_{\pi/2}^{\theta^\prime/2+\delta}
d\theta~\sin\theta\arccos
\left(\frac{\cos\delta-\cos\theta\cos(\theta^\prime/2)}
{\sin\theta\sin(\theta^\prime/2)}\right)\right],\\
\alpha_{b,2}(\delta)&=&\frac{2}{\pi}\int_{2\pi/3-\delta}^{\pi/2}
d\theta~\sin\theta\arccos\left(\frac{\cos\delta-\cos\theta\cos 2\pi/3}
{\sin\theta\sin 2\pi/3}\right),\\ \nonumber
\beta_{b,2}(\delta)&=&\frac{2}{\pi}\int_{2\pi/3-\delta}^{\pi/2}
d\theta~\sin\theta\arccos\left(\frac{\cos\delta-\cos\theta\cos 2\pi/3}
{\sin\theta\sin 2\pi/3}\right)+\\
&&\frac{2}{\pi}\int_{\pi/2}^{\pi/6+\delta}
d\theta~\sin\theta\arccos\left(\frac{\cos\delta-\cos\theta\cos \pi/6}
{\sin\theta\sin \pi/6}\right).
\eq

%%%%%%%%%%%%%%%%%%%%%%%%%%%%%%%%%%%%%%%%%%%%%%%%%%%%%%%%%%%%%%%%%%%%%%%%%%%%%%%
\bibliographystyle{apsrev}
\bibliography{shsdir2}

%%%%%%%%%%%%%%%%%%%%%%%%%%%%%%%%%%%%%%%%%%%%%%%%%%%%%%%%%%%%%%%%%%%%%%%%%%%%%%%
\newpage \centerline{\bf LIST OF FIGURES}

\begin{itemize}
\item[Fig. 1] Summary of the vector notation used to define the model
in the text.

\item[Fig. 2] Left panel: the adhesive patch model of Kern and Frenkel.\cite{Kern03}
Right panel: patch surface coverage $\chi_0$ as a function of
the patch angle $\delta$.

\item[Fig. 3] Adhesion requires simultaneous alignment of patches (dark shading)
on both spheres with the
vector between their centers. The spheres on the left do not adhere,
while those 
on the right do.

\item[Fig. 4] Configurations of three mutually bonded spheres, each possessing
a large single patch (dark shading).  The patch vectors point 
inward in the left panel and outward in the right panel.
The latter case is only possible for
$\delta>5\pi/6$.  Combinations of these arrangements are also
possible.

\item[Fig. 5] Dependence of the coefficients $\chi_i$ $(i=1,2,3)$
on $\delta$ for the one-patch model.

\item[Fig. 6] Configurations of three mutually bonded spheres, each possessing
two patches (dark shading).
On the left only one patch on each sphere is involved
in the bonds; on the right both patches on each sphere are involved.
The latter case is only possible for $\delta\ge\pi/3$).  Combinations
of these arrangements are also possible.

\item[Fig. 7] Dependence of the coefficients $\chi_i$ $(i=1,2,3)$
on $\delta$ for the two-patch model.

\item[Fig. 8] Radial distribution function for the one-patch model
($n=1$) at $\eta=0.4$ and $\tau=0.2$
within the C1 approximation and for various values of the adhesive
coverage $n\chi_0$.

\item[Fig. 9] Comparison between the radial distribution function from Monte
Carlo (MC) simulations and from the C0 and C1 approximations in the isotropic case ($n\chi_0=100\%$,
left-hand panel) and for an intermediate value of the single patch ($n=1$) case with $n\chi_0=80\%$ (right-hand panel). 
Both sets of calculations were performed at $\tau/(n\chi_0)^2=0.125$ and $\rho\sigma^3=0.35$ corresponding to $\eta=0.183\dots$.

\item[Fig. 10] Dependence on the adhesive coverage $n\chi_0$ of the binodal line calculated from
the modified Carnahan-Starling free energy of Eq.~(\ref{f-csm})
(left-hand panel)
and from the C1 approximation of Eq.~(\ref{f-c1}) (right-hand panel). The one-patch ($n=1$) and
two-patch ($n=2$) systems are compared at the same total coverage.

\item[Fig. 11] Dependence of the critical reduced temperature 
on the total adhesive coverage $n\chi_0$ for $n=1$ and $2$ patches, calculated from
the C1 approximation of Eq.~(\ref{f-c1}) and from MC.  The inset shows the
critical packing fraction.

\item[Fig. 12] Percolation thresholds for various total adhesive
  coverages as calculated from the C1 approximation (lines),
Eq.~(\ref{perc1}), and MC simulation (points). The
one-patch ($n=1$) and two-patch ($n=2$) cases are compared at the same total
coverage.

\item[Fig. 13] Average coordination numbers at the percolation threshold for
the one- and two-patch models at $60\%$ coverage, obtained
through MC.  The continuous line is the prediction from the integral
equation theory. The isotropic case is also reported for comparison.

\item[Fig. 14] Phase diagram in the C1 approximation, for various values of
the adhesive coverage $n\chi_0$. The one-patch ($n=1$) and
two-patch ($n=2$) models are compared at the same total coverage. 

\item[Fig. 15] Comparison of the C1 approximation with MC simulation (dots) for the
phase diagram of particles with two patches.  The MC isotropic phase diagram is
taken from Ref.~\onlinecite{Miller03}.

\item[Fig. 16] Dependence of the binodal line of patchy
adhesive spheres with a background adhesion
on the total surface coverage $n\chi_0$ of the patches: on the left as
calculated from the modified Carnahan-Starling free energy
of Eq.~(\ref{f-csm}); on the right as calculated from
the C1 approximation, Eq.~(\ref{f-c1}).

\item[Fig. 17] Dependence of the critical reduced temperature 
on the adhesive coverage in the patchy model with a uniform adhesive
background, calculated from the C1 approximation, Eq.~(\ref{f-c1}).  The
inset shows the behavior of the 
critical packing fraction.

\item[Fig. 18] Basic geometry for the calculation of $R_1(\delta)$
in Eq.~(\ref{Bop}).  The required solid
angle is the overlap of the two cones of width $\delta$
(the darkly shaded region in the sketch).
\end{itemize}

%%%%%%%%%%%%%%%%%%%%%%%%%%%%%%% Figures %%%%%%%%%%%%%%%%%%%%%%%%%%%%%%%%%%%%%%%%%%%%%%%%%%%%
\newpage
%%%%%%
%Fig1
%%%%%%
\newpage
%%%%%%%
%Fig2
%%%%%%%
\newpage
%%%%%%%
% Fig3
%%%%%%%
\newpage
%%%%%%%
% Fig4
%%%%%%%
\newpage 
%%%%%%%%%%%
% Fig5
%%%%%%%%%%%
\newpage 
%%%%%%%%%%
% Fig6
%%%%%%%%%%
\newpage
%%%%%%%%
% Fig7
%%%%%%%%
\newpage
%%%%%%%%%
% Fig8
%%%%%%%%%
\newpage
%%%%%%%
% Fig9
%%%%%%%%
\newpage
%%%%%%%%
% Fig10
%%%%%%%%
\newpage
%%%%%%%%%
% Fig11
%%%%%%%%%
\newpage
%%%%%%%%
% Fig12
%%%%%%%
\newpage
%%%%%%%%%
% Fig13
%%%%%%%%%
\newpage
%%%%%%%%
%Fig14
%%%%%%%%
\newpage
%%%%%%%%
% Fig15
%%%%%%%%
\newpage
%%%%%%%%%
% Fig16
%%%%%%%%%
\newpage
%%%%%%%%%%
% Fig17
%%%%%%%%%%
\newpage
%%%%%%%%%%%
% Fig18
%%%%%%%%%%%
%
%%%%%%%%%%%%%%%%%%%%%%%%%%%%%%%%%%%%%%%%%%%%%%%%%%%%%%%%%%%%%%%%%%%%%%%%%%%%%%%
%%%%%%%%%%%%%%%%%%%%%%%%%%%%%%%%%%%%%%%%%%%%%%%%%%%%%%%%%%%%%%%%%%%%%%%%%%%%%%%
%%%%%%%%%%%%%%%%%%%%%%%%%%%%%%%%%%%%%%%%%%%%%%%%%%%%%%%%%%%%%%%%%%%%%%%%%%%%%%%
\end{document}